\documentclass[12pt,preprint]{aastex}

\usepackage{epsfig}

\newcommand{\um}{$\mu$m~}
\newcommand{\ums}{$\mu$m}

\slugcomment{}

\shorttitle{}
\shortauthors{Farrah et al.}

\begin{document}

\title{Mid-Infrared Spectroscopy of Optically Faint Extragalactic 70$\mu$m Sources}

\author{
D.~Farrah\altaffilmark{1,2},
D. Weedman\altaffilmark{2},
C. J. Lonsdale\altaffilmark{3},
M. Polletta\altaffilmark{4},
M. Rowan-Robinson\altaffilmark{5},
J. Houck\altaffilmark{2},
H. E. Smith\altaffilmark{6}
}

\altaffiltext{1}{Astronomy Center, University of Sussex, Falmer, Brighton, BN1 9RH, UK}
\altaffiltext{2}{Astronomy Department, Cornell University, Ithaca, NY 14853, USA}
\altaffiltext{3}{ALMA Science Center, National Radio Astronomy Observatory, Edgemont Rd., Charlottesville, VA 22904, USA}
\altaffiltext{4}{INAF-ISAF Milano, via E. Bassini, Milan 20133, Italy}
\altaffiltext{5}{Astrophysics Group, Imperial College, London SW7 2BW, UK}
\altaffiltext{6}{Center for Astrophysics and Space Sciences, University of California at San Diego, La Jolla, CA 92093, USA; deceased 2007 August 16}

\begin{abstract}
We present mid-infrared spectra of sixteen optically faint sources with 70$\mu$m fluxes in the range 19mJy $<$ f$_{\nu}$(70\ums) $<$ 38mJy. The sample spans a redshift range of $0.35<z<1.9$, with most lying between $0.8<z<1.6$, and has infrared luminosities of $10^{12} - 10^{13}$L$_{\odot}$. Ten of 16 objects show prominent polycyclic aromatic hydrocarbon (PAH) emission features; four of 16 show weak PAHs and strong silicate absorption, and two objects have no discernable spectral features.  Compared to samples with f$_{\nu}$(24\ums) $>$ 10\,mJy, the 70\um sample has steeper IR continua and higher luminosities. The PAH dominated sources are among the brightest starbursts seen at any redshift, and reside in a redshift range where other selection methods turn up relatively few sources. The absorbed sources are at higher redshifts and have higher luminosities than the PAH dominated sources, and may show weaker luminosity evolution. We conclude that a 70$\mu$m selection extending to $\sim20$mJy, in combination with selections at mid-IR and far-IR wavelengths, is necessary to obtain a complete picture of the evolution of IR-luminous galaxies over $0<z<2$.
\end{abstract}


\section{Introduction} \label{intro}
Among the most important cosmological results of the last few decades was the discovery by the Cosmic Background Explorer (COBE) of a background radiation at infrared wavelengths \citep{pug,hau}. This background is comparable in intensity to the integrated optical light from the galaxies in the Hubble Deep Field, implying that the star formation rate density at $z\gtrsim1$ was more than an order of magnitude higher than locally, and that most of this star formation was obscured. Later surveys \citep{aus,dol,rr04} resolved the bulk of this background into a population of distant IR-luminous galaxies (LIRGs, $L_{ir}\gtrsim10^{11}L_{\odot}$) which undergo strong luminosity evolution with redshift ($(1+z)^{\sim4}$, e.g. \citealt{poz,lef}), reaching a comoving density at least 40 times greater at $z\sim1$ than in the local Universe \citep{elb}. Reviews of their properties can be found in \citet{san96} and \citet{lfs06}. 

Significant effort has been devoted to understanding the mechanisms driving the evolution of LIRGs. At low redshift, they are almost invariably mergers \citep{sur98,far01,bus02,vei02}, powered mainly by star formation \citep{gen98,rig99,ima07,veg08}, and reside in average density environments \citep{zau07}. LIRGs at high redshift also appear to be mainly starburst dominated merging systems \citep{far02b,cha03,sma04,tak06,bor06,val07,ber07,bri07}, though there are signs of differences compared to their low redshift counterparts; for example, weak X-ray emission \citep{fra03,wil03,iwa05}, different modes of star formation \citep{far08}, and a tendency to reside in overdense regions \citep{bla04,far06a,mag07}. 

Controversies remain, therefore, over how LIRGs may or may not evolve with redshift. Part of the reason for this is that an efficient census of LIRGs at $z\gtrsim0.5$ is difficult, as surveys conducted in a single IR band can miss a significant fraction of the LIRG population. For example, sub-mm surveys find large numbers of obscured starbursts at $z>1$ \citep{bar99,cha05,are07,cle08,dye08}, but few sources at $z<1$, and virtually no sources with `hot' dust \citep{bla02}. It is therefore essential that we survey for LIRGs in every IR band available to us and, having found them, systematically study them further. 

The {\it Spitzer} space telescope \citep{wer,soi08} has the capacity to revolutionize our understanding of LIRGs. The Infrared Array Camera (IRAC, \citealt{faz04}) and the Multiband Imaging Photometer for Spitzer (MIPS, \citealt{rie04}) imaging instruments, and the Infrared Spectrograph (IRS, \citealt{hou04}) all offer dramatic improvements in sensitivity and resolution over previously available facilities. In particular, the MIPS 70$\mu$m channel is ideal for studying high redshift LIRGs; at $z\lesssim2$ the rest-frame emission is always longward of 18$\mu$m, giving good sensitivity to both starbursts and AGN. The optically faint, high redshift 70$\mu$m Spitzer sources may be a prime example of the sources that previous surveys in the sub-millimeter or in the infrared (IR) have missed - in the sub-mm because the sources harbour dust that is too hot for current generation sub-mm cameras to see, and in the surveys of the Infrared Space Observatory (ISO) because of sensitivity limits at $>20\mu$m\footnote{e.g. the ELAIS ISO survey reached a limiting depth of $\sim100$mJy at 90$\mu$m \citep{rr04}, compared to $\sim20$mJy for large area surveys with MIPS at $70\mu$m)}.

In this paper, we use the IRS to observe a sample of 16 sources selected at 70$\mu$m using data from the Spitzer Wide Area Infrared Extragalactic survey (SWIRE, \citealt{lon,oli,dav06,tyrtt,wad07,ber07,shu08,sia08}) survey. Our aim is to explore the range in redshifts, luminosities and power sources seen in the optically faint 70$\mu$m population. We assume $\Omega=1$, $\Lambda=0.7$ and $H_{0}=70$ km s$^{-1}$ Mpc$^{-1}$. Luminosities are quoted in units of ergs s$^{-1}$ or of bolometric solar luminosities, where $L_{\odot} = 3.826\times10^{33}$ ergs s$^{-1}$.

\section{Methods} \label{obsanl}

\subsection{Sample Selection} 
The sources are selected from the SWIRE Lockman Hole field, which covers 10.6 square degrees and reaches 5$\sigma$ depths of 4.2$\mu$Jy at 3.6$\mu$m, 7.5$\mu$Jy at 4.5$\mu$m, 46$\mu$Jy at 5.8$\mu$m, 47$\mu$Jy at 8.0$\mu$m, 209$\mu$Jy at 24$\mu$m, 18mJy at 70$\mu$m, and 108mJy at 160$\mu$m.  The primary selection criterion for our sample is a confident detection at 70$\mu$m, so we first rejected all sources fainter than 19mJy at 70$\mu$m. To ensure that we could obtain mid-infrared IRS spectra with reasonable signal-to-noise, we also constrained the sources to have f$_{\nu}$(24\ums) $>$ 0.9\,mJy, although  $\gtrsim 90\%$ of sources with $f_{70}>19$mJy also satisfy $f_{24}>0.9$mJy. This resulted in a parent sample of 1250 sources.  From this, we selected optically faint sources by taking all sources (12 in total) with $r$-band magnitudes fainter than $m_{r}=23$, and including an additional four sources with $r$-band magnitudes in the range $20 < m_{r} < 23$, for a total of 16 objects.

\subsection{Observations} 
All 16 objects were observed as part of Spitzer program 30364 with the first order of the short-low module (SL1; 7.4$\mu$m - 14.5$\mu$m, slit size $3.7\arcsec\times57\arcsec$ with 1.8\arcsec\ pix$^{-1}$, R$\sim60-127$), and the second order of the long-low module (LL2; 14.0$\mu$m - 21.3$\mu$m, slit size $10.5\arcsec\times168\arcsec$ with 5.1\arcsec\ pix$^{-1}$, R$\sim57-126$). Eight of these objects were additionally observed with long-low order 1 (LL1, 19.5$\mu$m - 38.0$\mu$m). The targets were placed in the center of each slit using the blue peak-up array. Each target was observed with an individual ramp time of 60s in SL, and 120s in LL, with the number of ramps determined by the targets  24$\mu$m flux density. Details are given in Table \ref{obslog}.

The data were processed through the {\it Spitzer} Science Center's pipeline software (version 15.3), which performs standard tasks such as ramp fitting and dark current subtraction, and produces Basic Calibrated Data (BCD) frames. Starting with these frames, we removed rogue pixels using the {\em irsclean}\footnote{This tool is available from the SSC website: http://ssc.spitzer.caltech.edu} tool and campaign-based pixel masks. The individual frames at each nod position were then combined into a single image using the SMART software package \citep{hig}. Sky background was removed from each image by subtracting the image for the same object taken with the other nod position (i.e. `nod-nod' sky subtraction). One-dimensional spectra were then extracted from the images using the SPICE software package using `optimal' extraction and default parameters. This procedure results in separate spectra for each nod and for each order. The spectra for each nod were inspected; features present in only one nod were treated as artifacts and removed. The two nod positions were then combined. The first and last 4 pixels on the edge of each order, corresponding to regions of decreased sensitivity on the array, were then removed, and the spectra in different orders merged, to give the final spectrum for each object.

\section{Results} \label{results}
\begin{figure}
\begin{minipage}{180mm}
\includegraphics[angle=90,width=70mm]{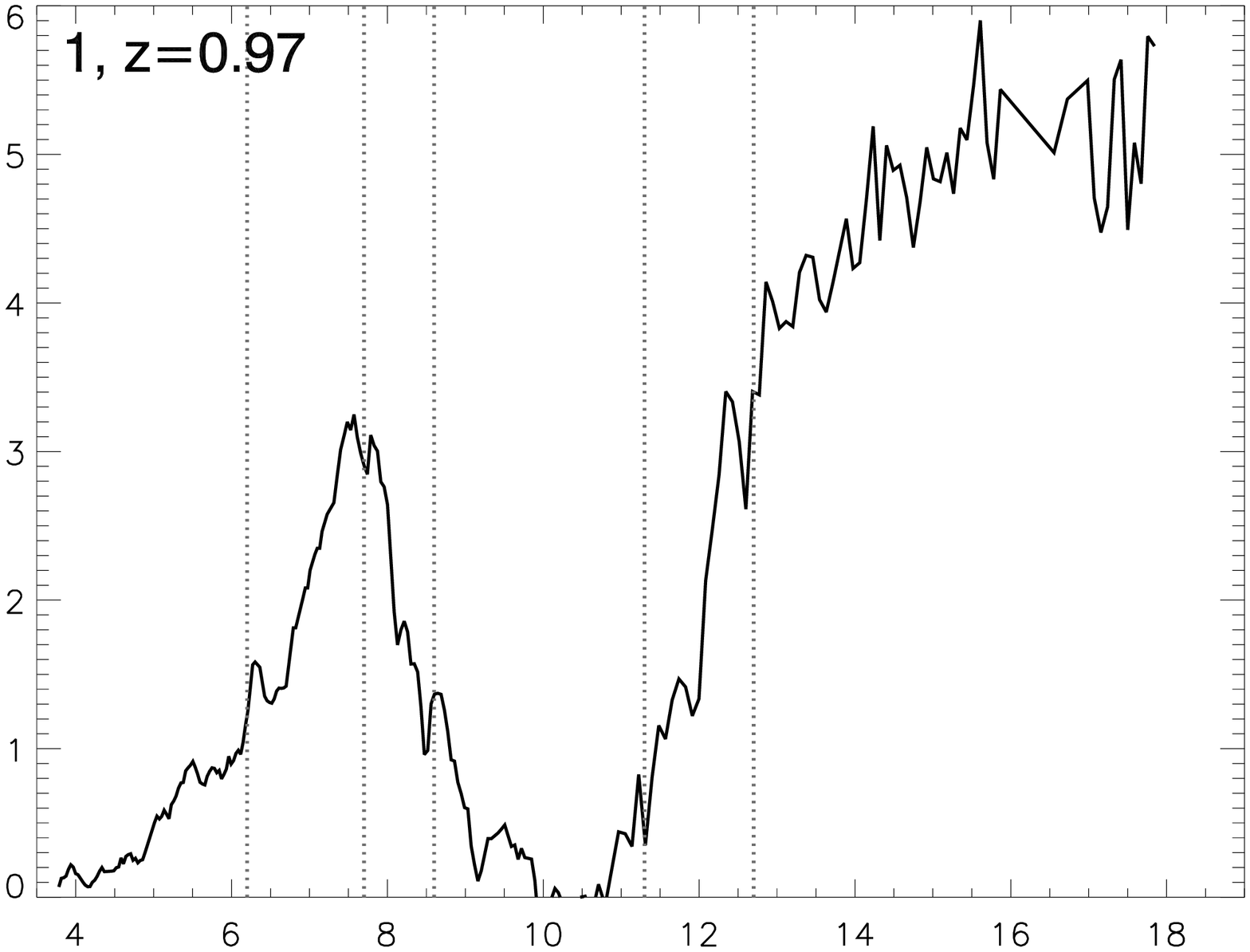}
\includegraphics[angle=90,width=70mm]{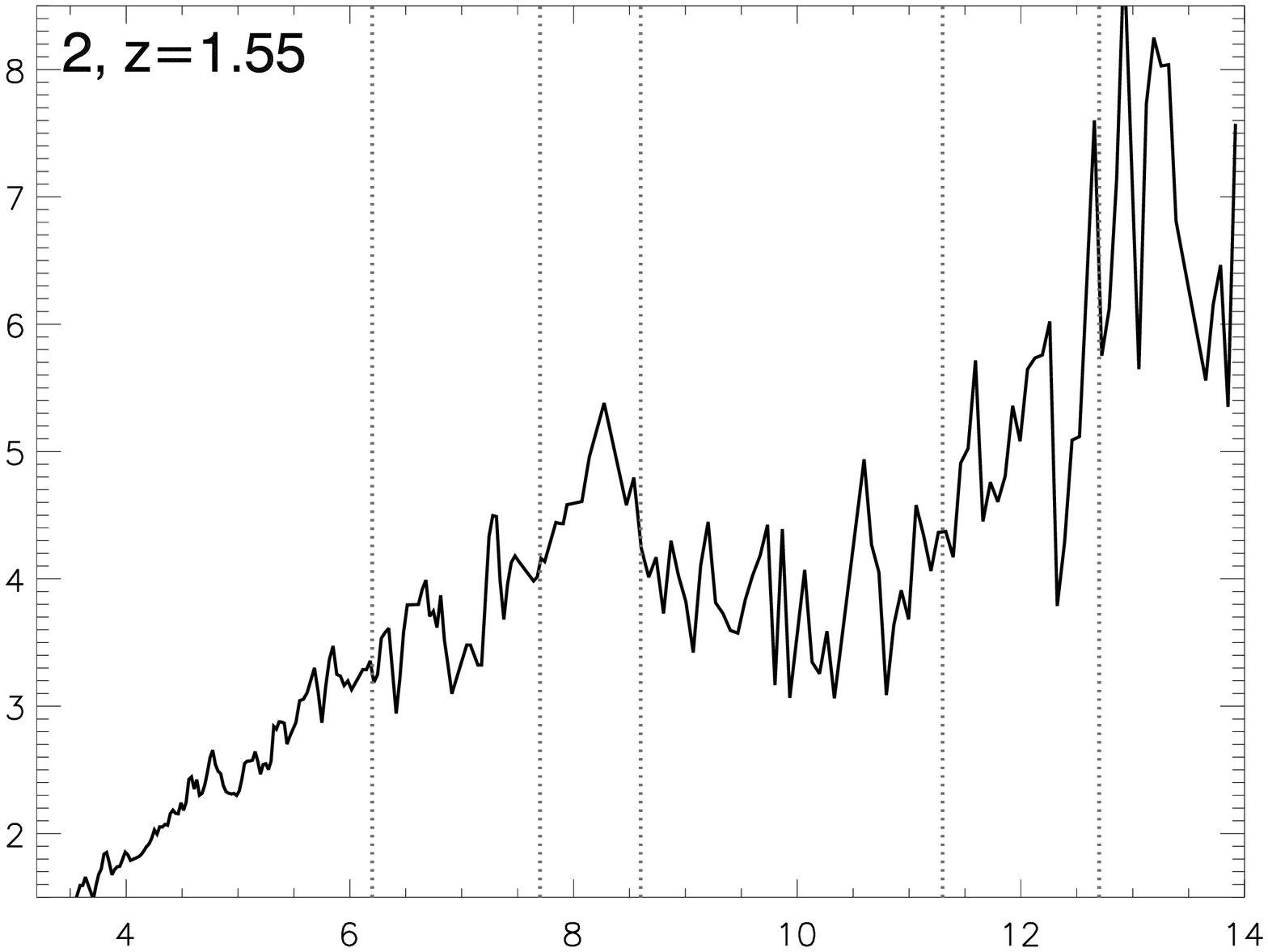}
\end{minipage}
\begin{minipage}{180mm}
\includegraphics[angle=90,width=70mm]{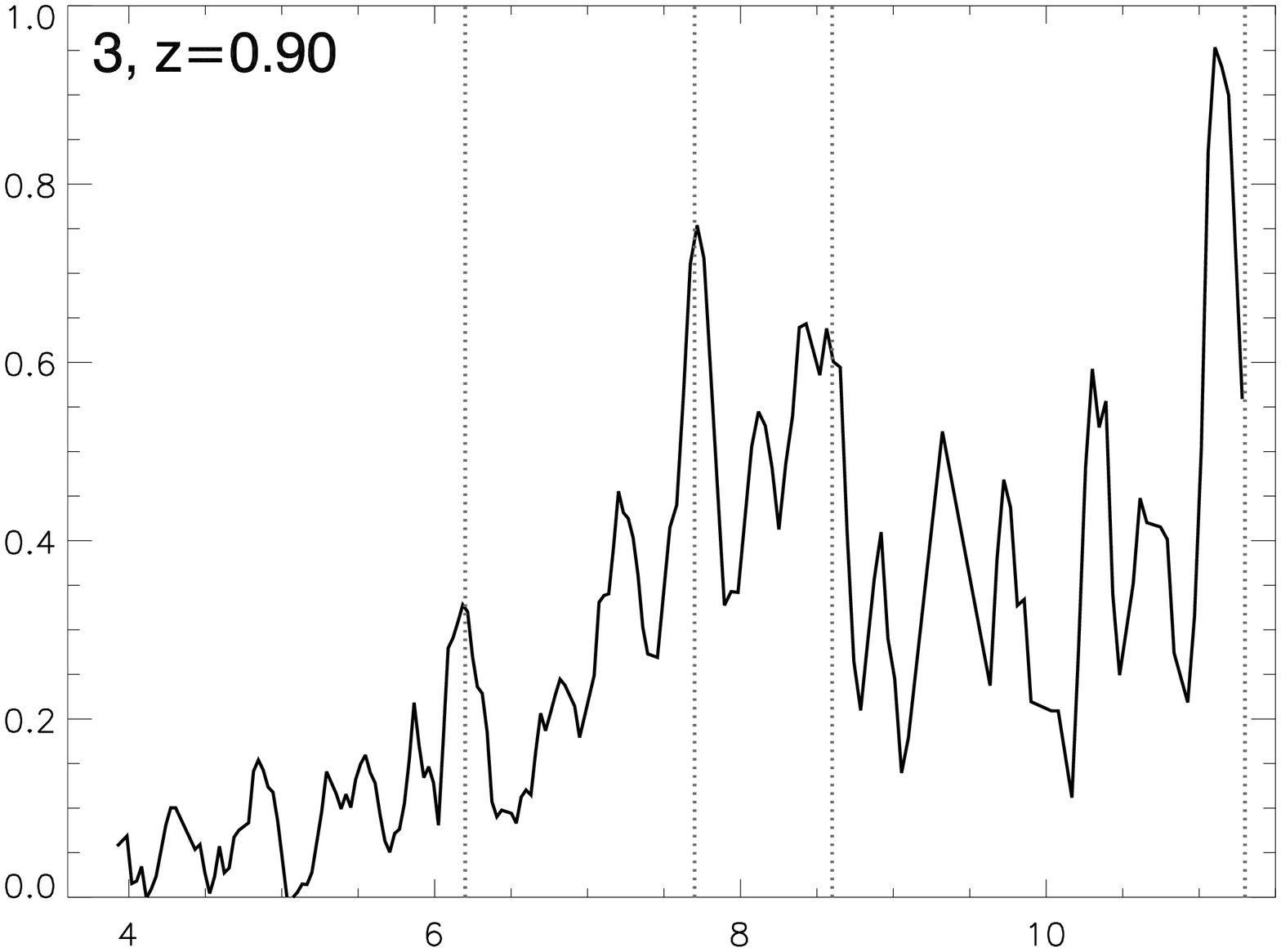}
\includegraphics[angle=90,width=70mm]{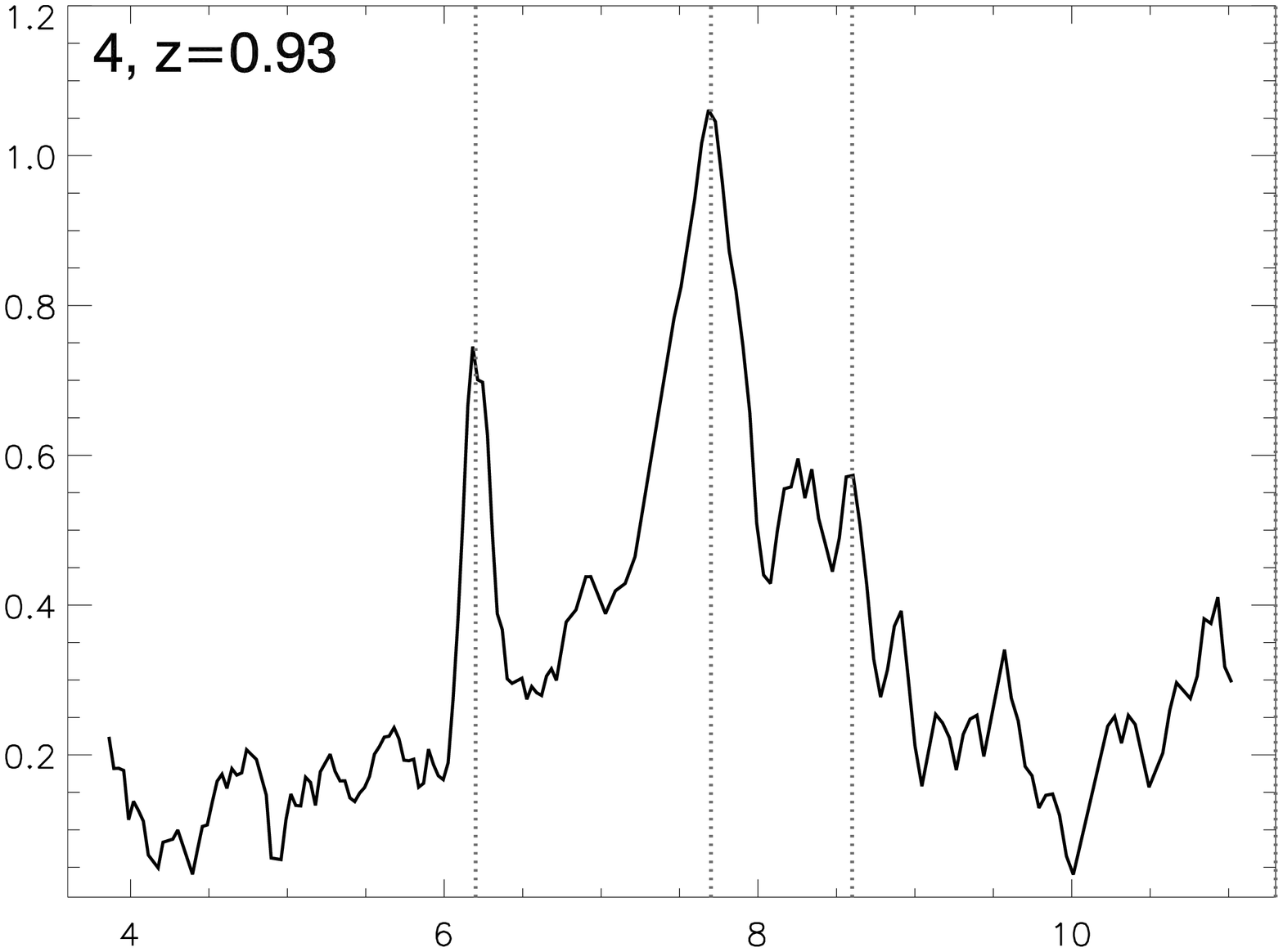}
\end{minipage}
\begin{minipage}{180mm}
\includegraphics[angle=90,width=70mm]{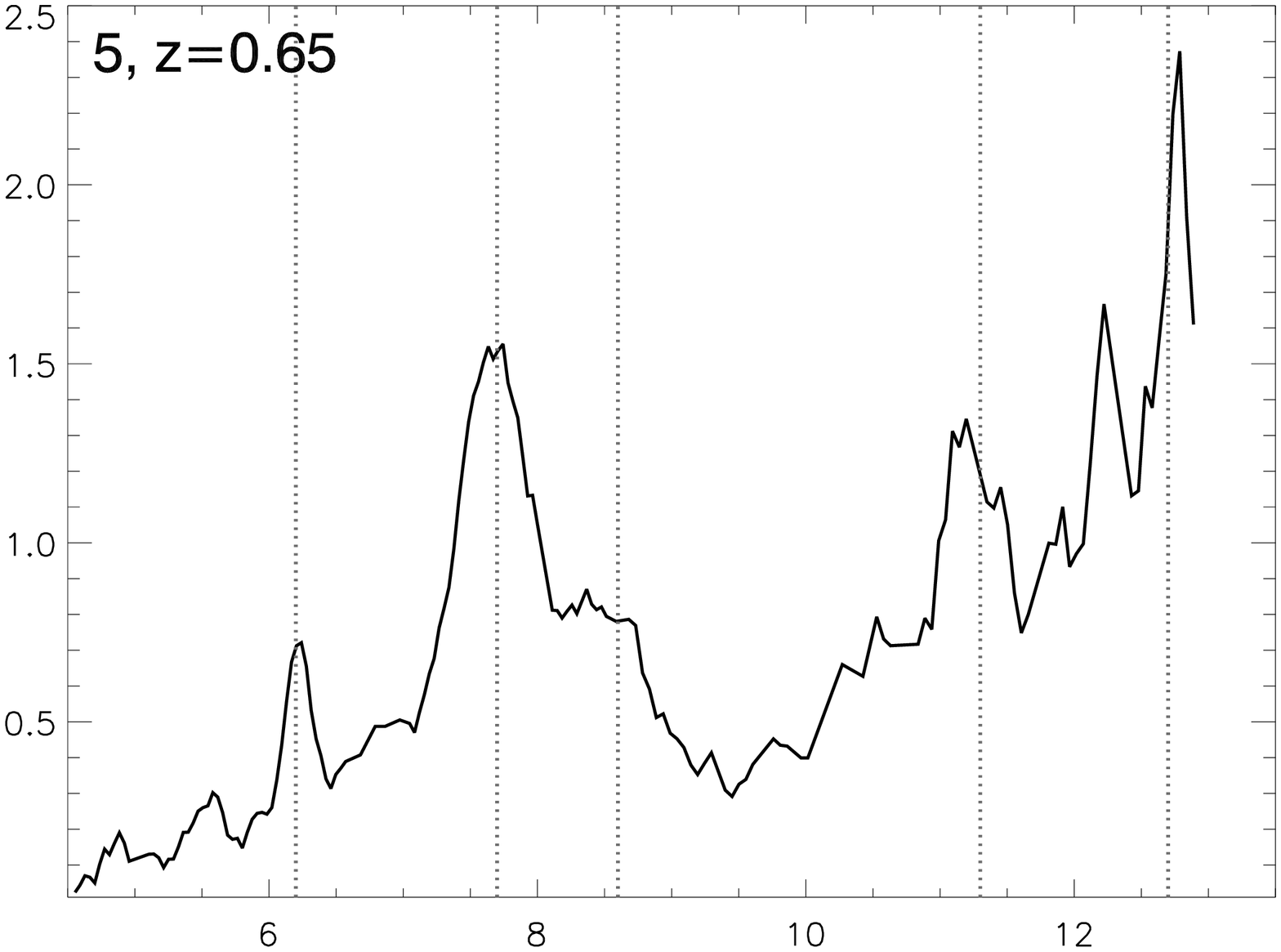}
\includegraphics[angle=90,width=70mm]{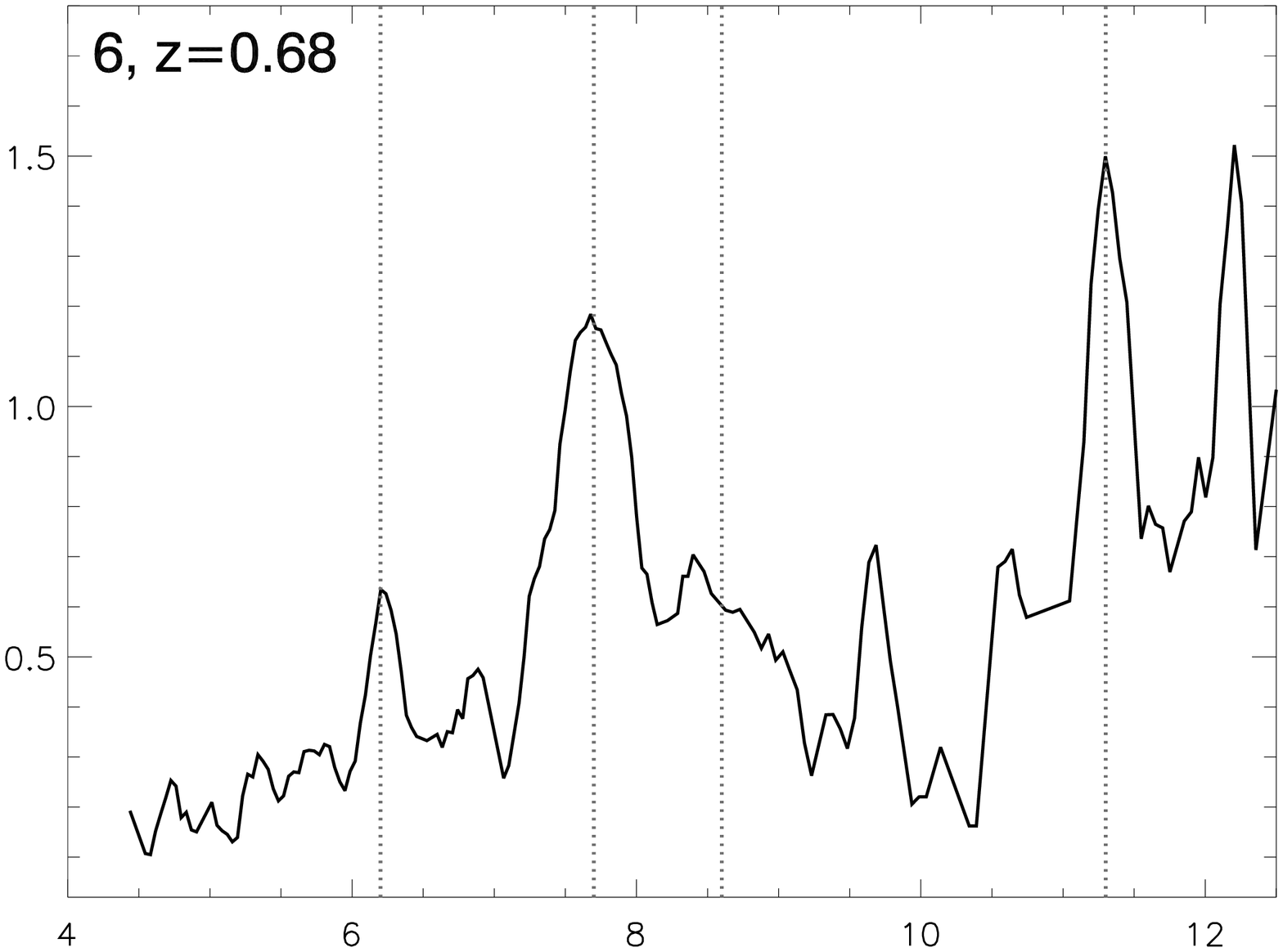}
\end{minipage}
\begin{minipage}{180mm}
\includegraphics[angle=90,width=70mm]{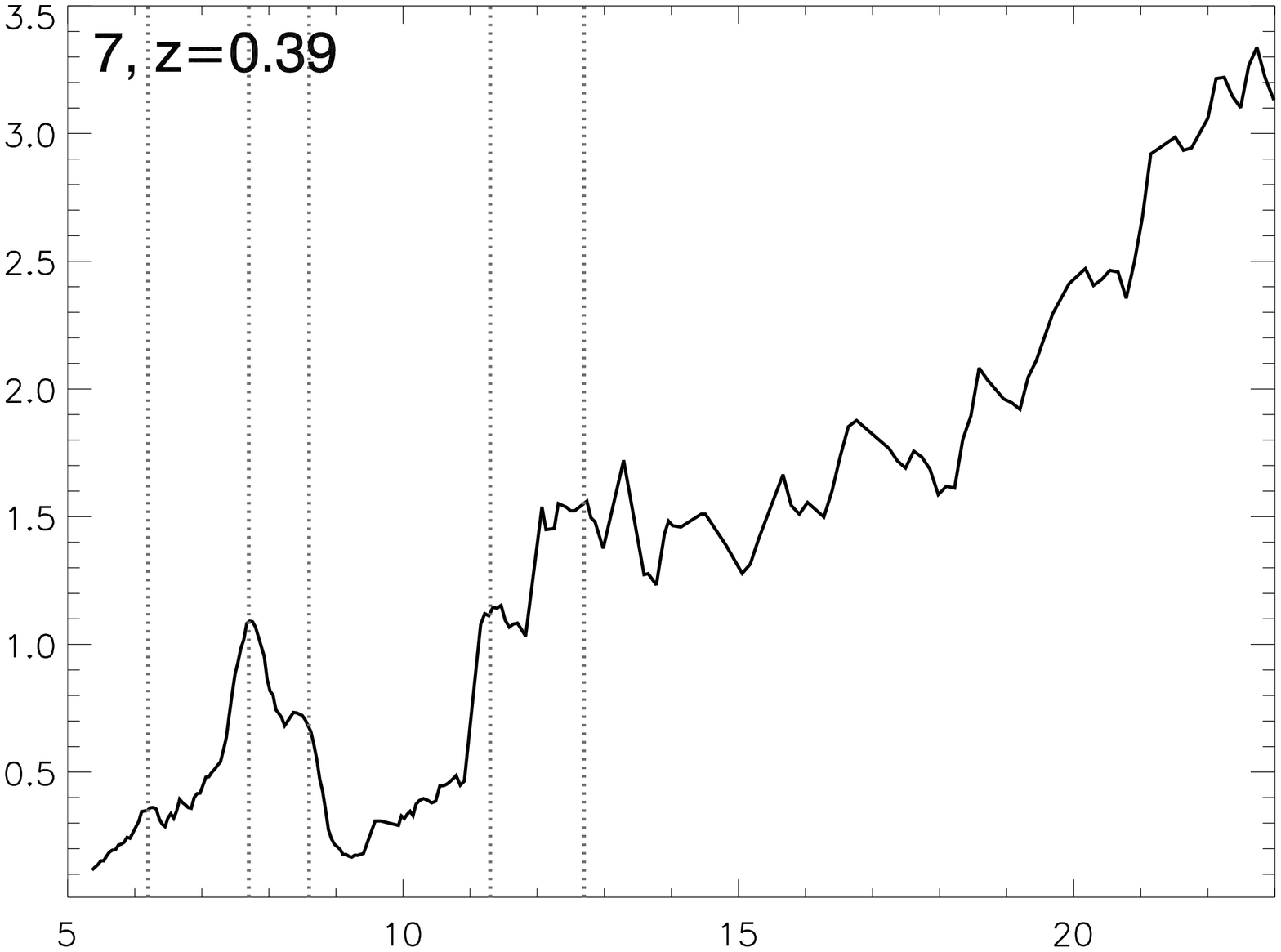}
\includegraphics[angle=90,width=70mm]{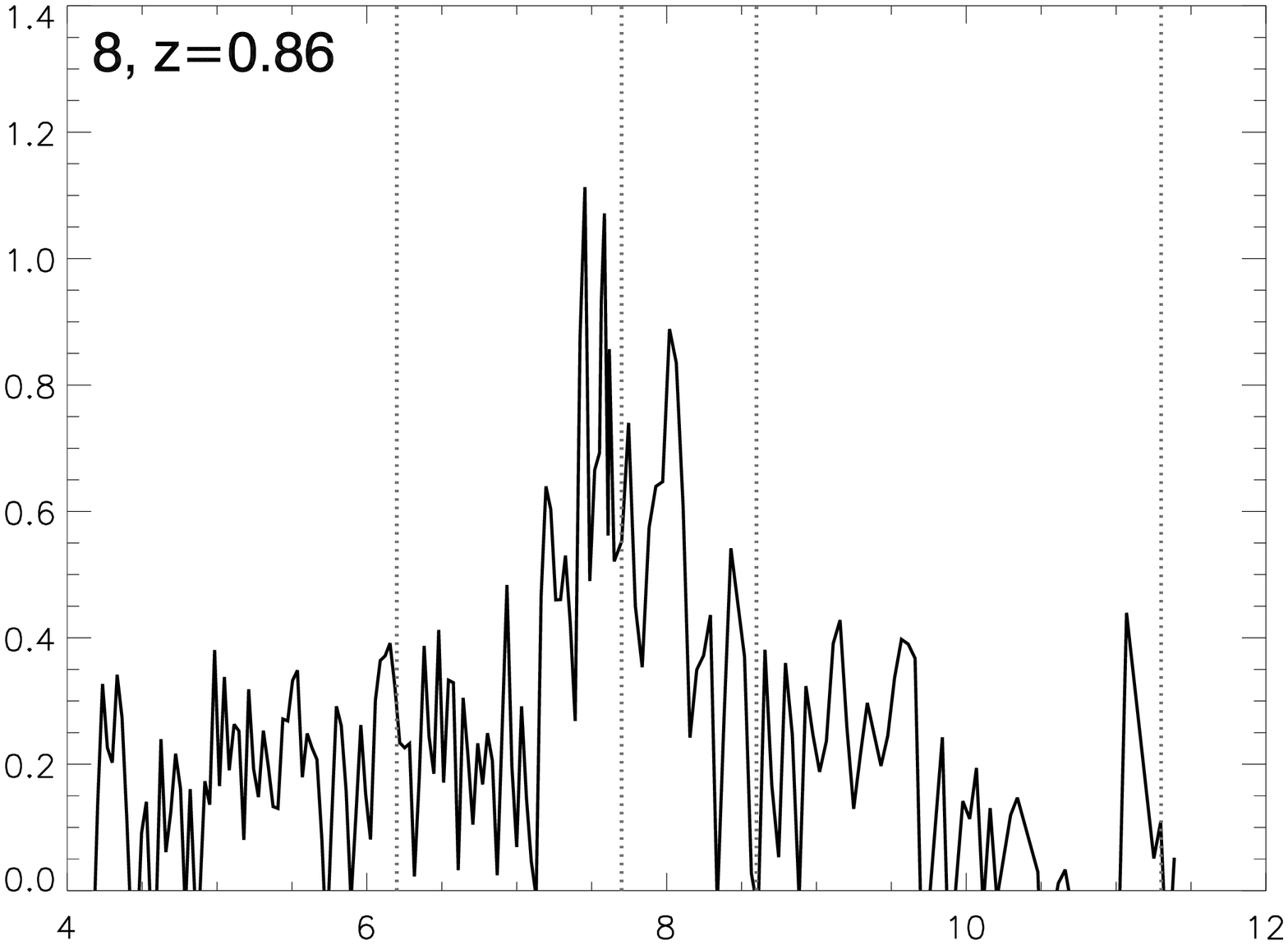}
\end{minipage}
\caption{Individual spectra of sources with wavelength in $\mu$m and flux density in mJy, plotted in the rest-frame using the IRS redshifts in 
Table \ref{sample}. The numbers in the top left of each panel are the ID number in Table \ref{obslog} followed by the redshift. The vertical dotted lines mark the 
6.2$\mu$m, 7.7$\mu$m, 8.6$\mu$m, 11.2$\mu$m and 12.7$\mu$m PAH features.
\label{spectraa}}
\end{figure}

\begin{figure}
\begin{minipage}{180mm}
\includegraphics[angle=90,width=70mm]{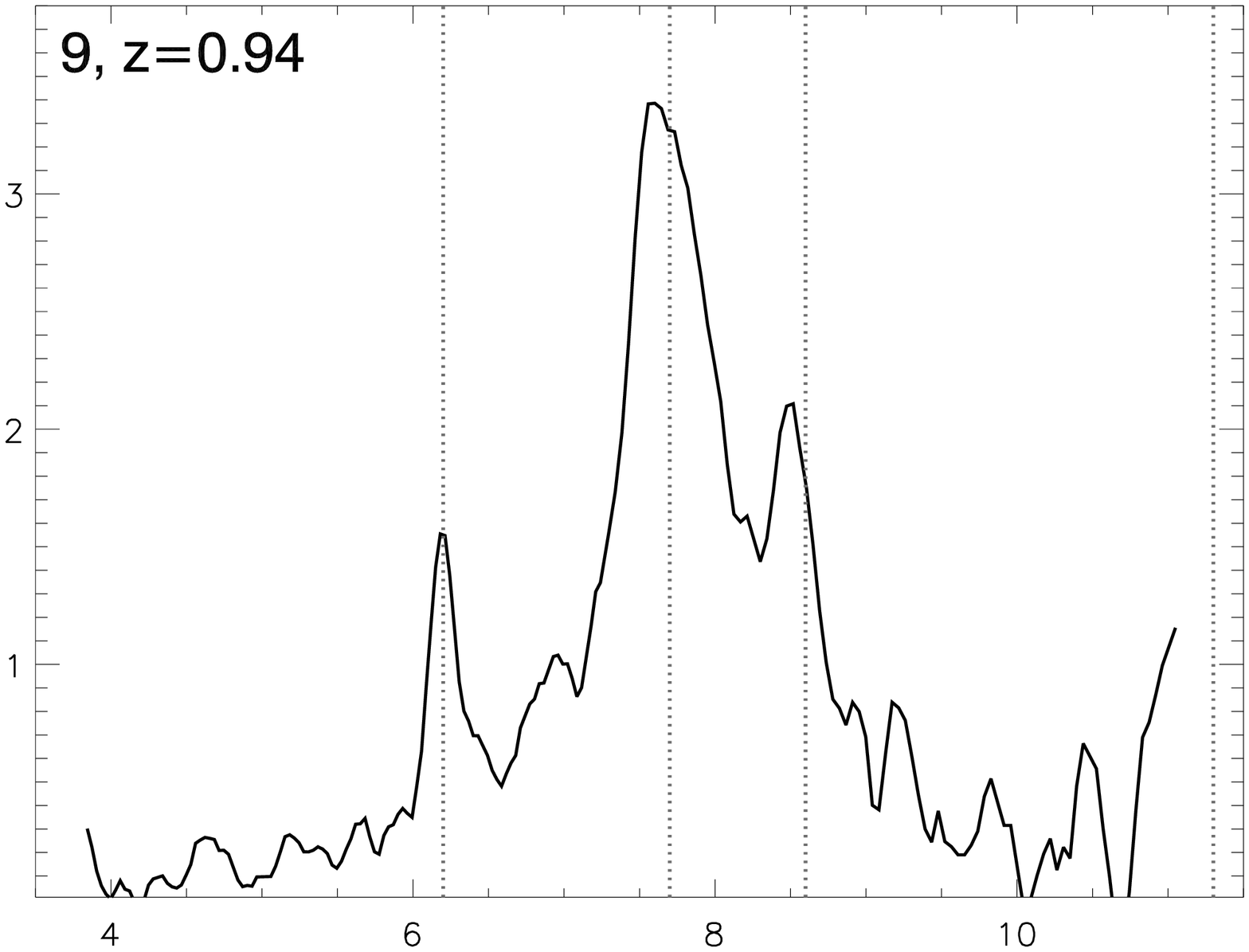}
\includegraphics[angle=90,width=70mm]{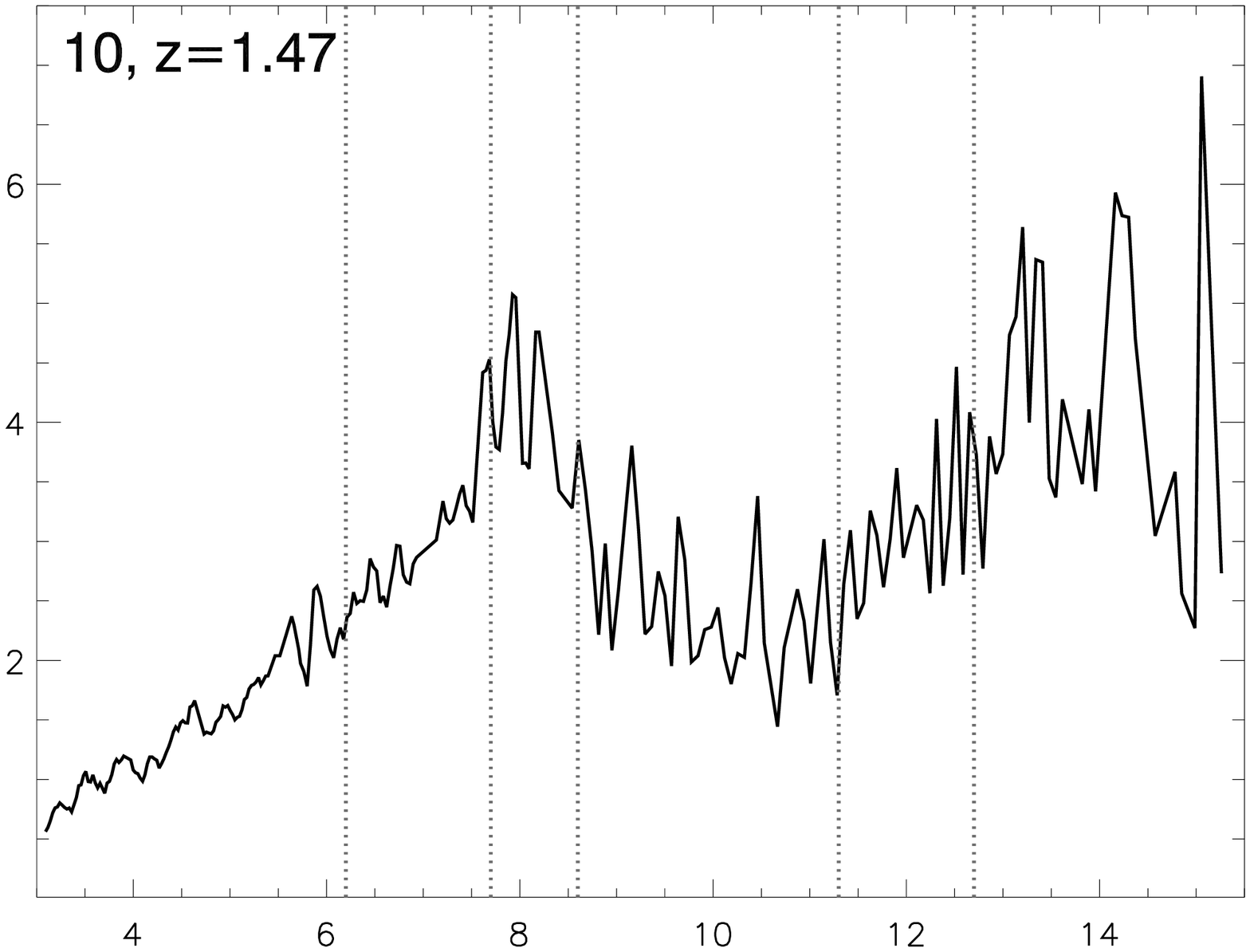}
\end{minipage}
\begin{minipage}{180mm}
\includegraphics[angle=90,width=70mm]{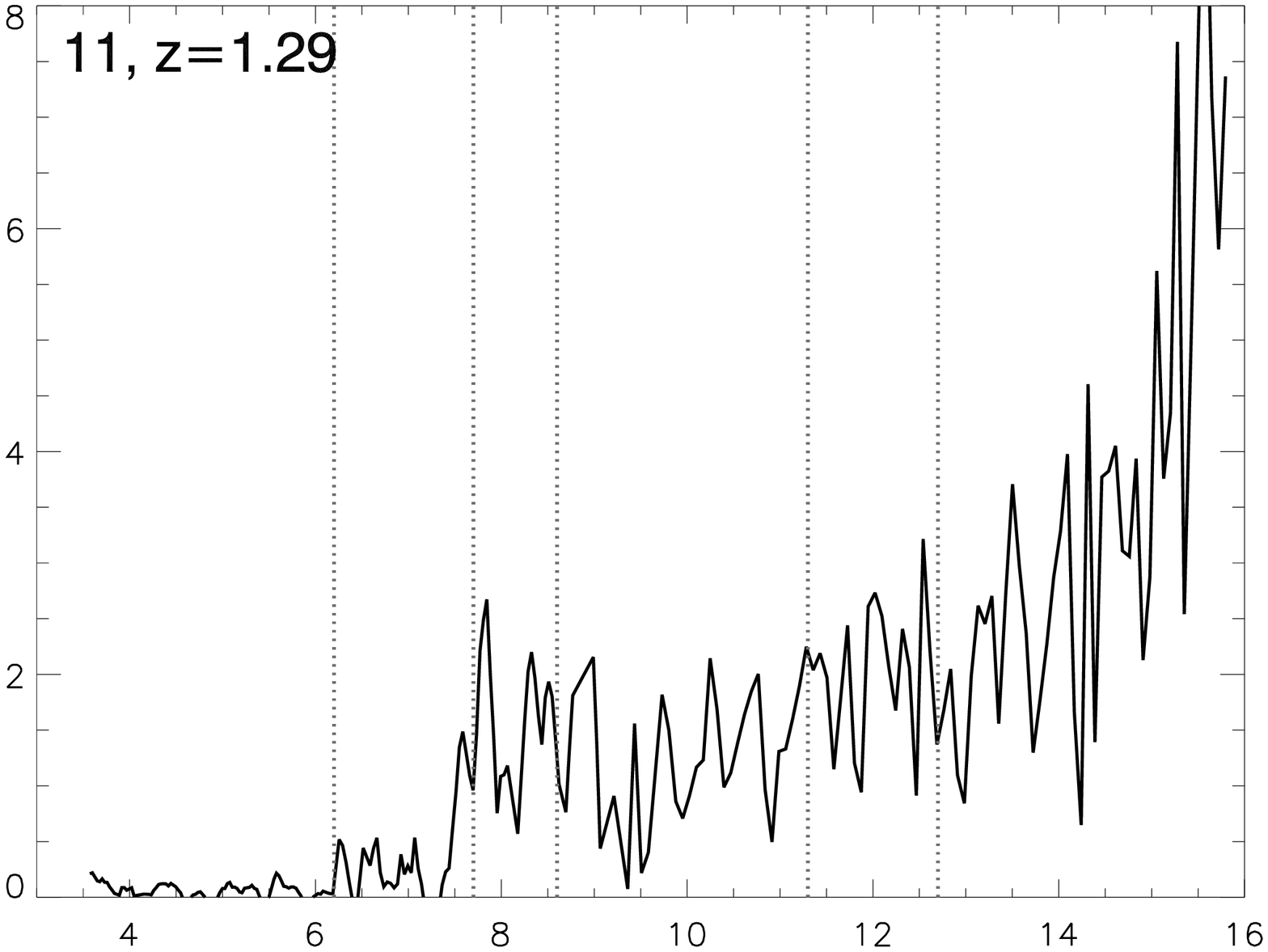}
\includegraphics[angle=90,width=70mm]{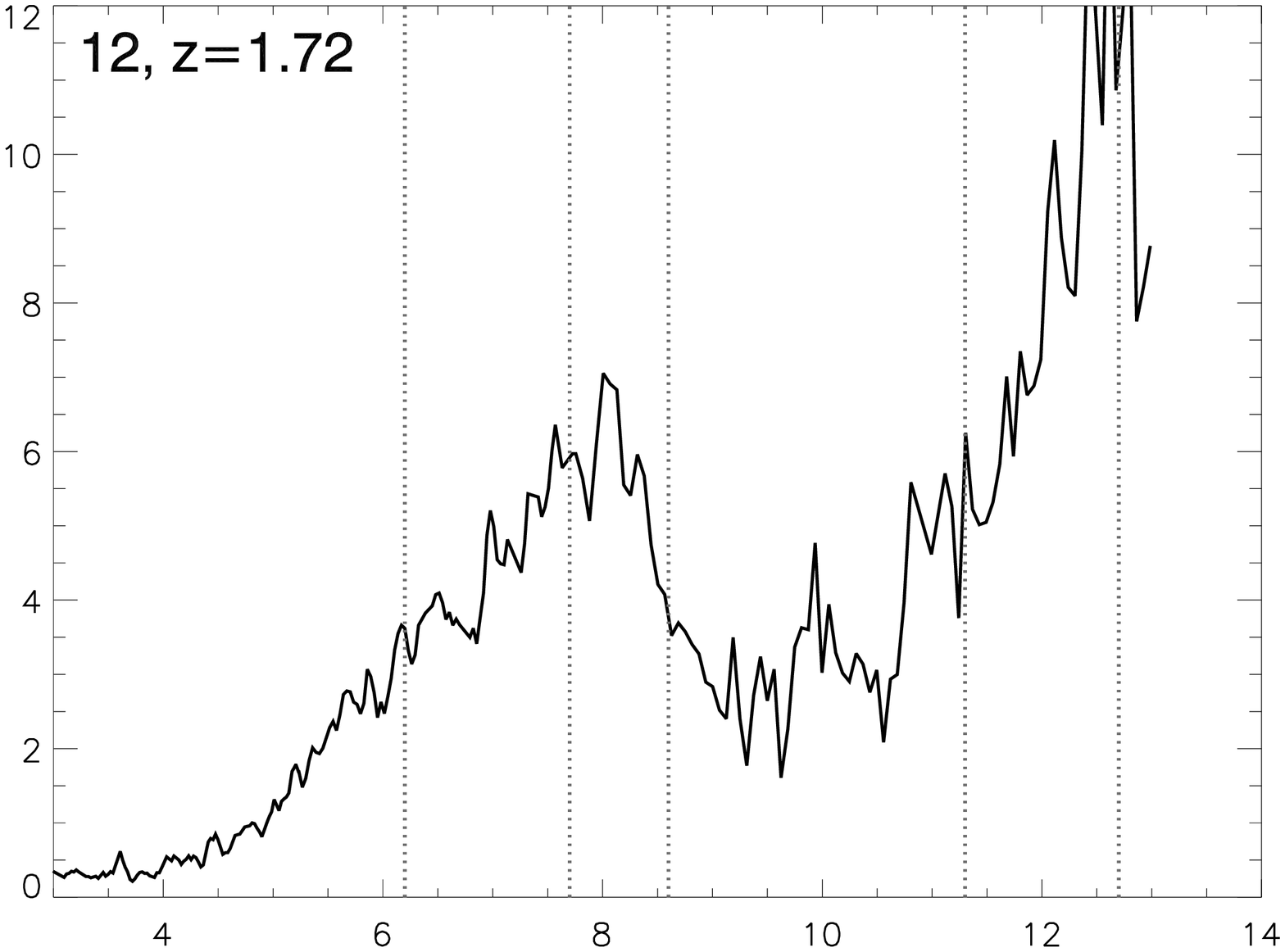}
\end{minipage}
\begin{minipage}{180mm}
\includegraphics[angle=90,width=70mm]{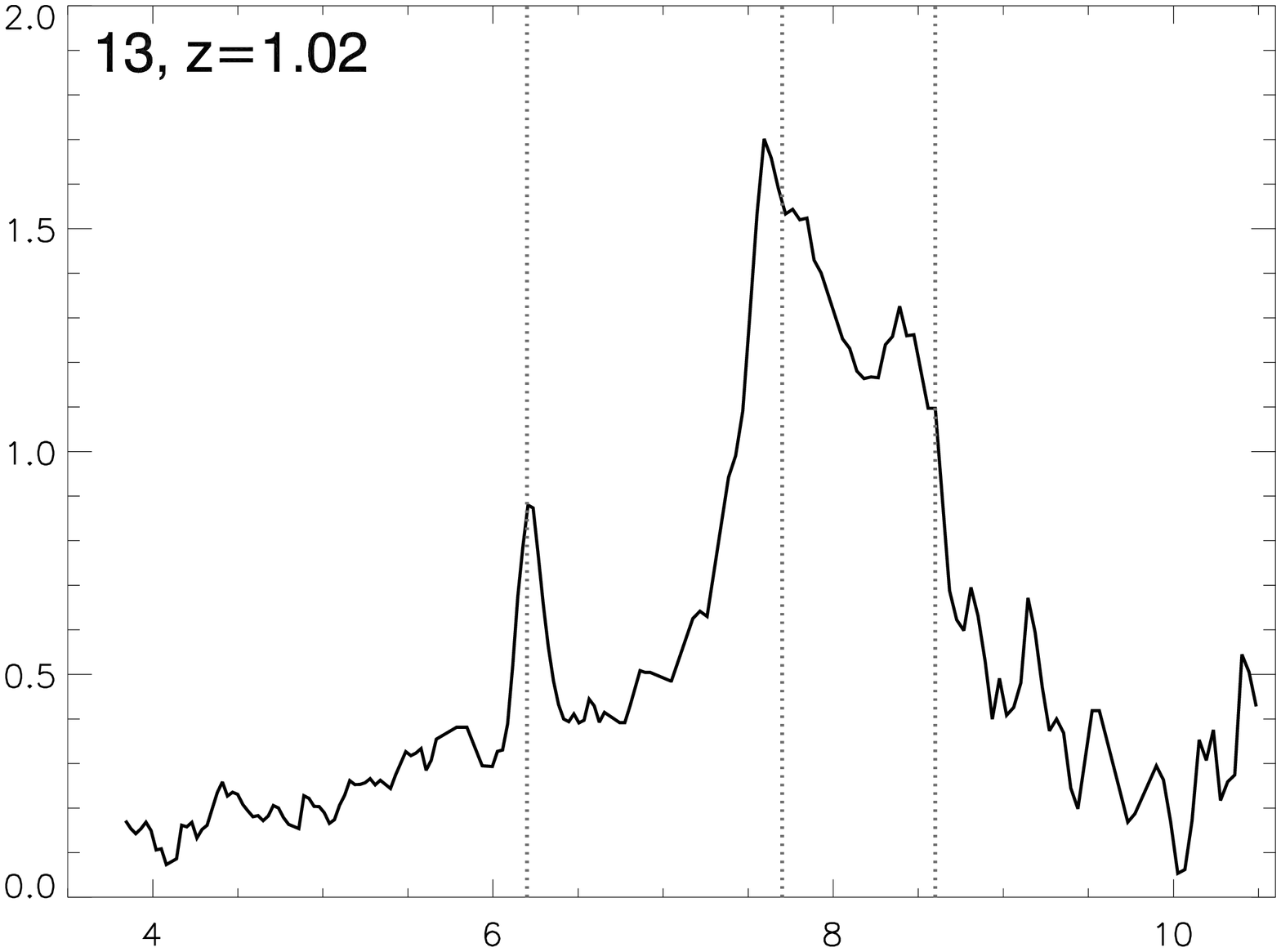}
\includegraphics[angle=90,width=70mm]{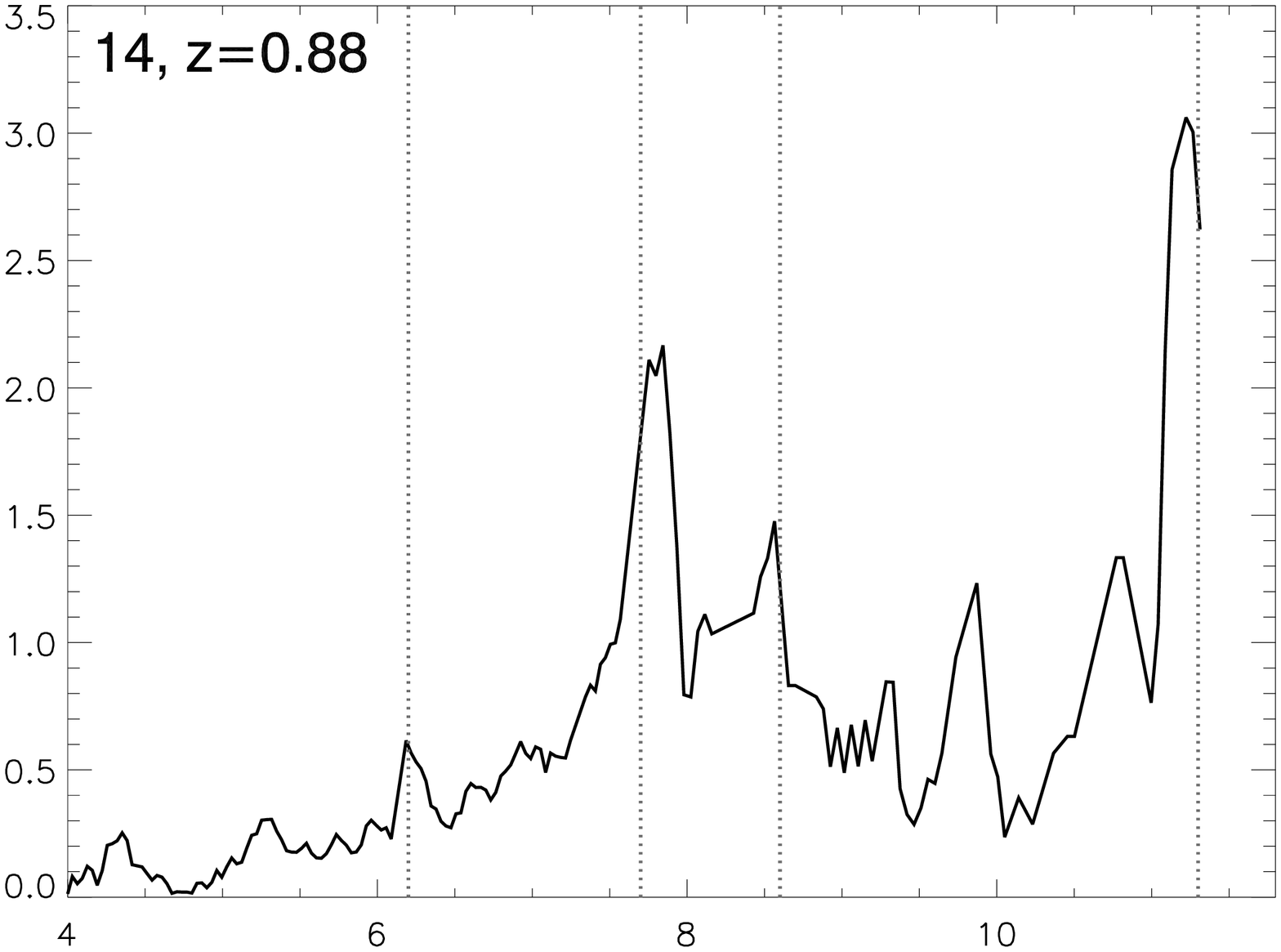}
\end{minipage}
\begin{minipage}{180mm}
\includegraphics[angle=90,width=70mm]{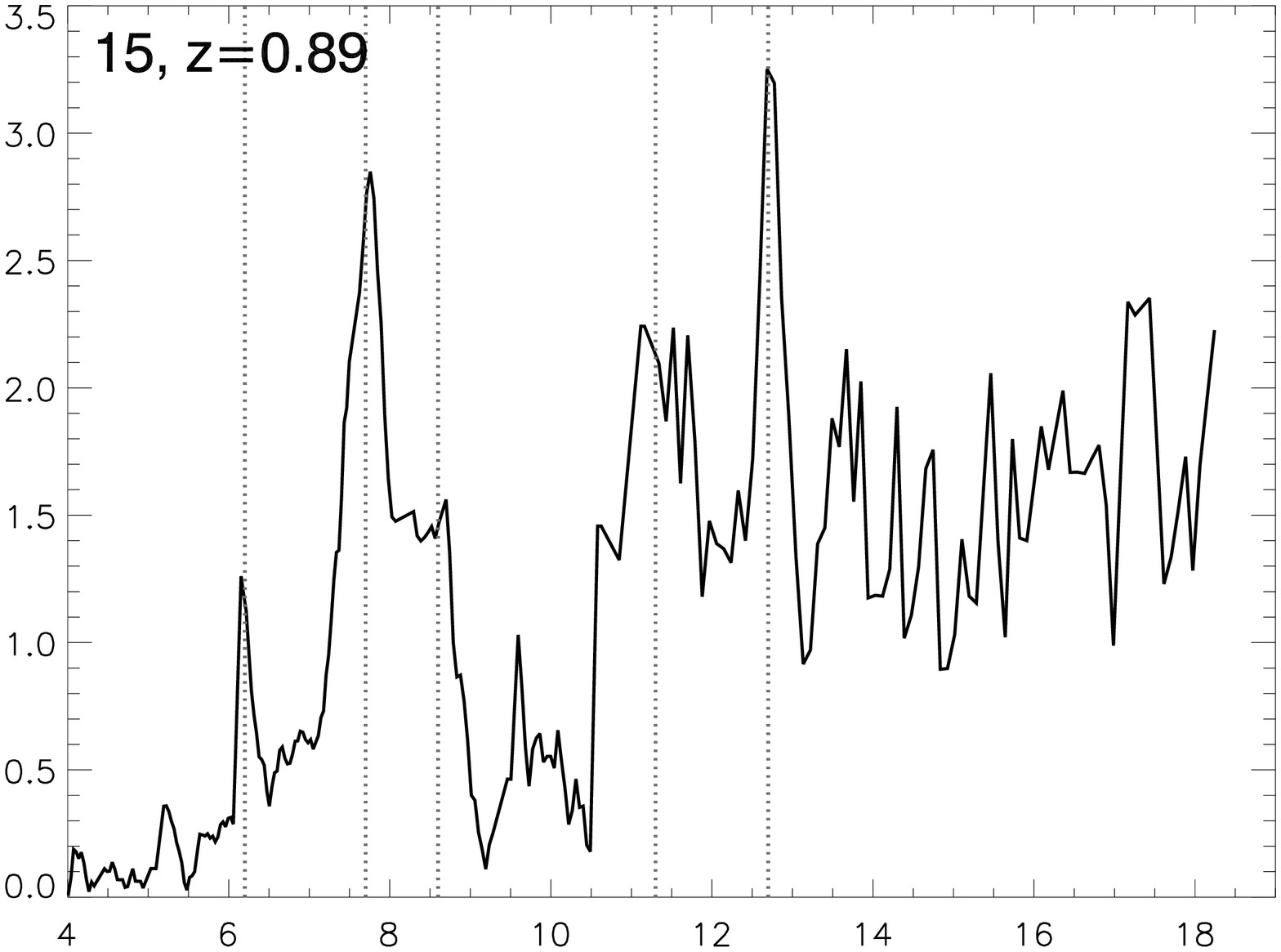}
\includegraphics[angle=90,width=70mm]{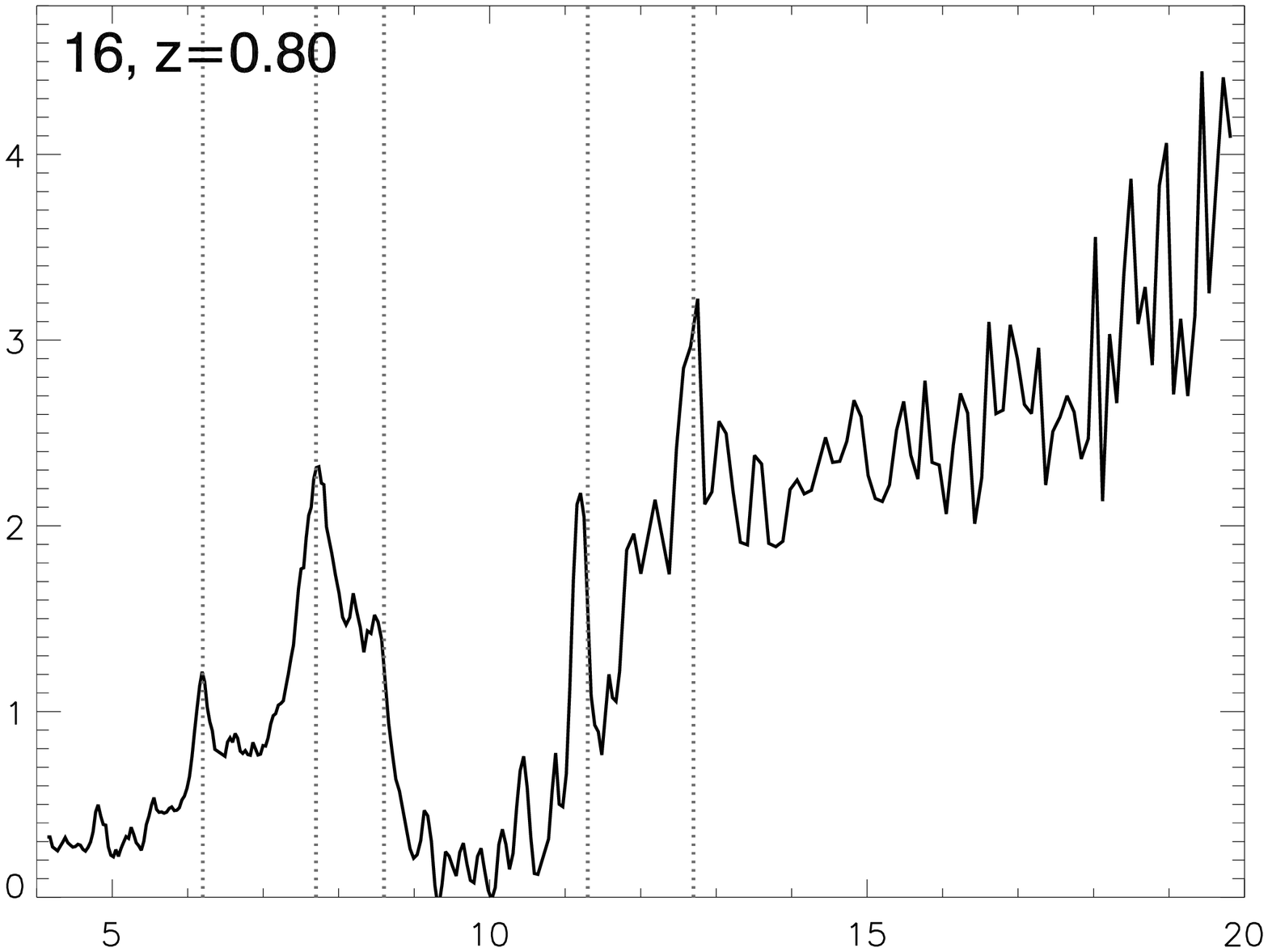}
\end{minipage}
\caption{Individual spectra, continued from Figure \ref{spectraa}
\label{spectrab}}
\end{figure}

The spectra are presented in Figures \ref{spectraa} and \ref{spectrab}. Redshifts and fluxes are given in Table \ref{sample}, and spectral measurements are given in Table \ref{pahsils}.

\subsection{Redshifts} 
We derive spectroscopic redshifts from broad emission features at 6.2$\mu$m, 7.7$\mu$m, 8.6$\mu$m, 11.2$\mu$m and 12.7$\mu$m, attributed to bending and stretching modes in neutral and ionized Polycyclic Aromatic Hydrocarbon (PAH) molecules (the 12.7$\mu$m feature also contains a contribution from the [NeII]$\lambda$12.81 fine-structure line), and/or a broad absorption feature centered at 9.7$\mu$m arising from large silicate dust grains. For eleven sources (1, 3-7, 9, 13-16), an unambiguous redshift can be determined from the PAH features; the uncertainty on these redshifts is governed by the variations in PAH peak wavelengths seen in local galaxies and is of order $\Delta z=0.02$. In three cases (2, 10, 12), the PAHs are weak, and the redshifts are derived from a prominent silicate absorption feature. In these cases, the redshifts have a larger error, $\Delta z\simeq 0.2$, but should still be reliable. Finally, in two cases (8 \& 11), no spectral features can be unambiguously identified; we derive tentative redshifts based on what are plausibly PAH or silicate features. In these cases, the error on the redshift is large, $\Delta z\simeq 0.4$, and the redshifts should be treated with caution. 

\subsection{Spectral properties}\label{specresults}
We measure the properties of the PAH features using two methods. For the 6.2$\mu$m and 11.2$\mu$m PAH features, we compute fluxes and equivalent widths (EWs) by integrating the flux above a spline interpolated local continuum fit (for a description of the method see \citealt{bra06} and \citealt{spo07}). The errors on the EWs are large because the continua of our sample are only weakly detected. Using this method means, however, that our EWs can be compared directly to those of local LIRGs as measured by Spitzer \citep{wee05,bra06,arm07,spo07}. For the 7.7$\mu$m PAH feature, we do not attempt to measure EWs because of the uncertainties in determining a continuum baseline underneath this broad and complex feature.  Instead, we measure only the flux density at the peak of the 7.7$\mu$m feature, as in \citet{hou07} and \citet{wee08}.  Due to the low S/N and restricted wavelength range of the spectra, we cannot correct for water ice and/or aliphatic hydrocarbon absorption, although the effect of this lack of correction is likely to be insignificant.

For the four objects with clear detections of the 6.2$\mu$m and 11.2$\mu$m PAH features, we derive star formation rates via the formula from \citet{far07b}; this yields SFRs of between 100 and 300 M$_{\odot}$ per year.  We derive star formation rates in Table 3 for all sources using the formula in \citet{hou07}:

\begin{equation}\label{sfreq}
log(SFR) [M_{\odot} yr^{-1}] = log(\nu L_{\nu} (7.7\mu m)) - 42.57
\end{equation}

\noindent For the four objects for which both formulae can be used, we obtain consistent results within the uncertainties, which are of order $50\%$. 

We measured the strengths of the silicate features, $S_{sil}$, via:

\begin{equation}
S_{sil} = ln\left(  \frac{F_{obs}(9.7\mu m)}{F_{cont}(9.7\mu m)} \right)\label{silstrength}
\end{equation}

\noindent where $F_{obs}$ is the observed flux density at rest-frame 9.7$\mu$m, and $F_{cont}$ is the underlying continuum flux density at rest-frame 9.7$\mu$m deduced from a spline fit to the continuum on either side. A description of this method can be found in \citet{spo07}, \citet{lev07} and \citet{sir07}.

Other than the PAH and silicate features, we see weak but clear detections of the H$_{2}$S(3) line at 9.66$\mu$m in two objects (6 and 14), but no other spectral features.

\subsection{SED Fitting}\label{sedfits}
We measure the IR luminosities of the sample by fitting the IR photometry simultaneously with the library of model spectral energy distributions (SEDs) for the emission from a starburst \citep{efs00} and an AGN \citep{rr95}, following the methods in \citet{far03}. The fits were good, with $\chi^{2}\lesssim2$ in all cases, and the observed-frame 70$\mu$m flux gave good constraints on the IR luminosities (see also discussion in \citet{rr05}). The photometry is too limited, however, to provide meaningful constraints on the starburst and AGN fractions, so we only present the derived total IR luminosities, and not the SED fits. The luminosities are presented in Table \ref{pahsils}. All objects have IR luminosities exceeding $\sim$10$^{11.5}$L$_{\odot}$, with most lying in the range 10$^{12} - 10^{13}$L$_{\odot}$, making them ULIRGs.

\section{Discussion}\label{disc}

\subsection{Redshifts and luminosities}
\begin{figure}
\begin{minipage}{180mm}
\includegraphics[angle=90,width=85mm]{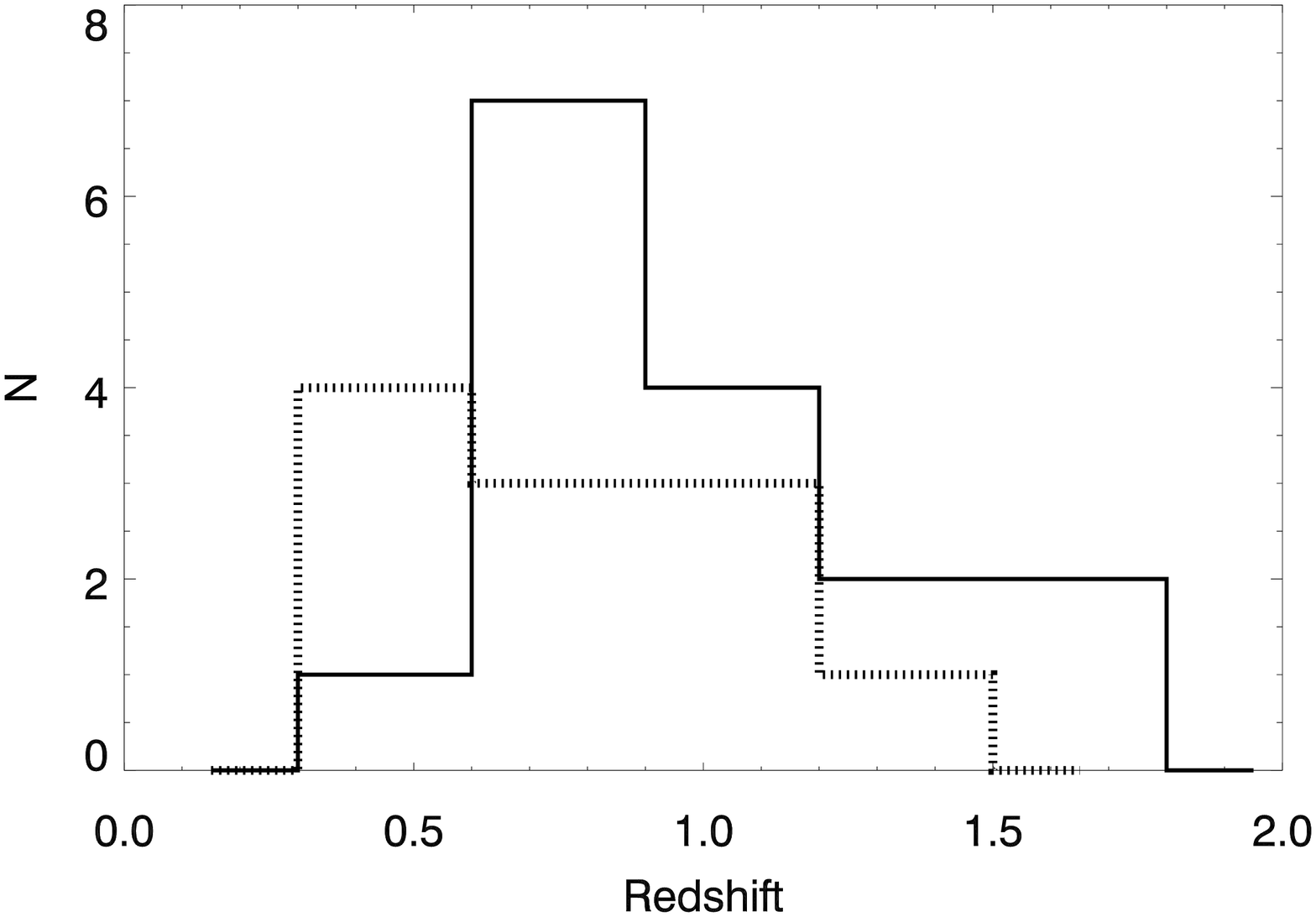}
\includegraphics[angle=90,width=85mm]{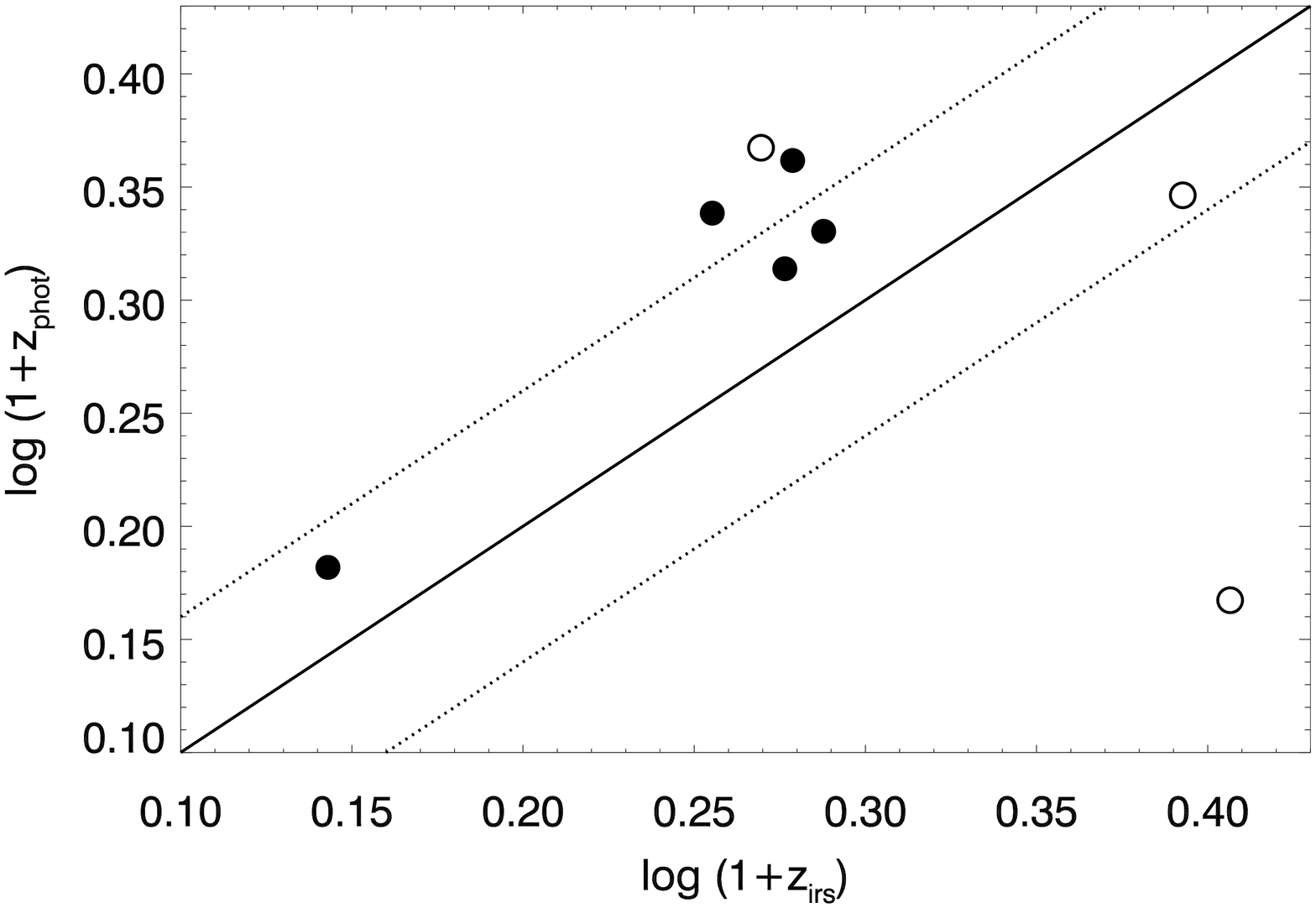}
\end{minipage}
\caption{{\it Left panel:} Redshift histogram, using the IRS redshifts. The solid line is the distribution for our sample. The dotted line is the distribution for the sample of \citet{brn08}. {\it Right panel:} Comparison between the IRS redshifts and the photometric redshifts from \citet{rr07} (Table \ref{sample}). The dotted lines denote a deviation of 0.06 in log(1+z), the boundary defined as a `catastrophic failure' by \citet{rr07}. Open symbols (objects 2, 8 and 10) have an unreliable IRS and/or photometric redshift. 
\label{zphotzspec}}
\end{figure}

The redshift distribution for our sample is shown in the left panel of Figure \ref{zphotzspec}. All objects lie in the range $0.35<z<1.8$. There  is a peak at z $\simeq0.9$, a long tail up to z $\simeq2$, and a shorter tail down to z $\simeq0.4$. 

The redshift range $0.5 \lesssim z \lesssim 1.5$ has been a difficult one in which to select ULIRGs because their distances make them faint at observed-frame 10-100$\mu$m, and the $k$-correction that makes ULIRGs bright at observed-frame 200-1000$\mu$m does not become strong enough for current sub-mm cameras until $z\gtrsim1.5$. Moreover, this is the redshift range where the evolution in the ULIRG luminosity function is thought to be strongest. Therefore, our simple selection method, consisting of little more than a minimum 70$\mu$m flux and an optical cut, should prove invaluable in studying the cosmological evolution of ULIRGs from future wide-area surveys. It seems likely that it is the optical cut that is resulting in our sources mostly lying in the $0.5 \lesssim z \lesssim 1.5$ range, as other spectroscopic surveys of purely 70$\mu$m selected sources find a lower median redshift. For example, \citet{huy07} present redshifts for 143 sources selected solely on the basis of 70$\mu$m flux but to a much fainter limit of $f_{70}>2$mJy, and find a median redshift of 0.64, with $79\%$ of the sources lying at $z<1$ and about half the sources at $z<0.5$. 

Given the difficulties in obtaining spectroscopic redshifts for distant ULIRGs, it is useful to assess the accuracy of photometric redshifts for this type of source. In the right panel of Figure \ref{zphotzspec}, we compare the spectroscopic redshifts to the photometric redshifts derived by \citet{rr07} for those eight objects where the photometric redshift code produced a formally acceptable fit ($\chi^{2}<10$, see discussion in \citet{rr07}). The photometric redshifts are reasonably good, given the faintness and high redshifts of the sample. Seven of 8 objects lie within or close to the `catastrophic failure' boundary of  $\log(1+z_{phot})=\log(1+z_{spec})\pm0.06$. Only in one case is there a clear mismatch between $z_{phot}$ and $z_{spec}$, and the photometric redshift for this object is unreliable, as it is based on limited data.

\subsection{Comparisons to other samples}

\subsubsection{Local ULIRGs}
\begin{figure}
\begin{minipage}{180mm}
\includegraphics[angle=90,width=170mm]{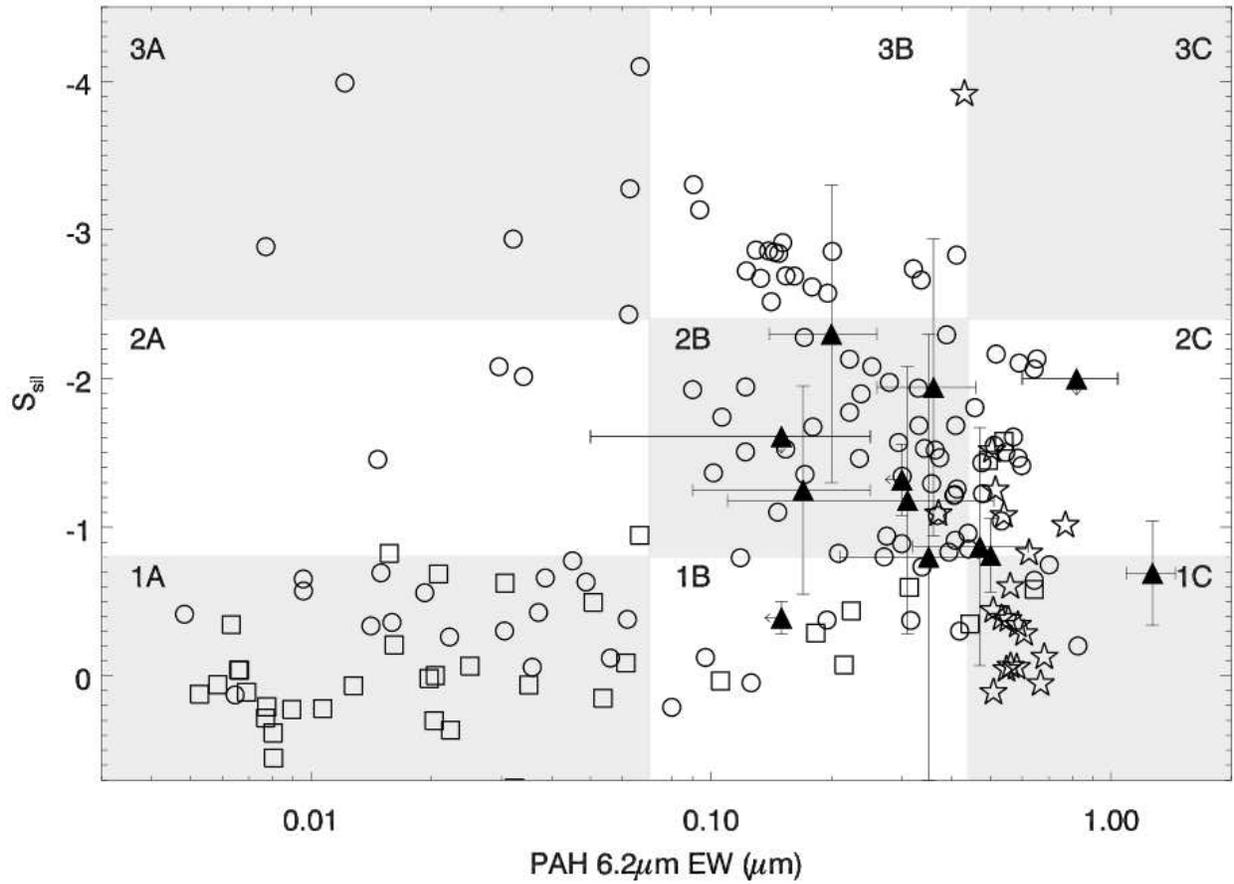}
\end{minipage}
\caption{The `Fork' diagram of \citet{spo07}. {\it Triangles}: Our 70$\mu$m sources.  {\it Circles}: Local ULIRGs.  {\it Stars, squares}: Classical starbursts \citep{bra06} and AGN \citep{wee05} with IR luminosities in the range 10$^{10} - 10^{11}$L$_{\odot}$, respectively. 
\label{fork}}
\end{figure}

We first compare our mid-IR spectra to those of local ULIRGs. Our sample is faint at 24$\mu$m, so detailed spectral diagnostics are not possible. We therefore employ a simple comparison using the `Fork' diagram of \citet{spo07}, shown in Figure \ref{fork}. Our 70$\mu$m sample has similar PAH and silicate absorption properties to the 1B/1C/2B/2C classes from \citet{spo07}. This identifies our sample with moderately obscured star-forming sources, but not with the most heavily absorbed sources or those that contain an unabsorbed or silicate-emitting AGN without PAH emission. 

\subsubsection{70$\mu$m selected samples}\label{o70comp}
The 70$\mu$m population has been studied relatively little with the IRS. The only other published study is that of \citet{brn08}, who select 11 sources with $f_{70}>32$mJy and an $r$ band magnitude fainter than $m_{r}=20$. Thus, the two samples make for interesting comparisons; our sample is 1-2 magnitudes fainter in the optical and $\sim1.6$ times fainter at 70$\mu$m. 

The IR luminosities of both samples are comparable, with both lying mostly in the 10$^{12} - 10^{13}$L$_{\odot}$ range. The fraction of sources with prominent PAHs is also similar with 7 of 11 PAH dominated sources in \citet{brn08} and 10 of 16 in ours. There are, however, two areas where there are differences, albeit with the caveat of small sample sizes. The first is redshift distribution. The redshift distribution for the Brand et al sample is overplotted in the left panel of Figure \ref{zphotzspec}. The Brand et al sample has a less pronounced peak, a broader distribution with more sources over $0.5<z<1$ and fewer sources at $z>1.2$, than does our sample. The second difference is the distribution of spectral types with redshift. The Brand et al sample shows no discernible separation of spectral type with redshift whereas our sample shows that all of the sources with strong PAHs, irrespective of the presence of silicate absorption, lie towards the lower end of the redshift range, while the strongly absorbed sources with negligible PAHS are all at the upper end\footnote{This is not a result of the 70$\mu$m selection shifting specific spectral features in and out of the bandpass, as there are no prominent features that would lie at observed-frame $\sim70\mu$m at the redshifts of our sample}. 

Both differences probably arise due to our sample reaching fainter 70$\mu$m fluxes than the sample of Brand et al. In principle, surveys to fainter 70$\mu$m fluxes should include higher redshift, more luminous sources (see \S\ref{comp24micron}). Moreover, a 70$\mu$m selection should result in sensitivity to different effective dust temperatures at different redshifts; at $z\simeq0.7$, $70\mu$m observations sample rest-frame 40$\mu$m, while at $z\simeq1.5$, they sample rest-frame 28$\mu$m, so at $z\sim1.5$ we are sensitive to sources with $\sim30$K hotter dust than at $z\sim0.7$. Therefore, higher redshift sources in a 70$\mu$m selected sample are more likely to be absorbed, AGN-like sources with weak PAH features, which is what we see in our sample. The absence of this trend in the Brand et al sample suggests that optically faint sources with $f_{70}\gtrsim 30$mJy are mainly ULIRGs at moderate redshift, but that at 70$\mu$m fluxes below 30mJy we start to see significant numbers of heavily absorbed sources, with large masses of hot dust, at $z>1$. Interestingly, a similar situation is seen at longer wavelengths; sub-mm surveys are adept at picking up sources with large masses of cold dust, but radio surveys in the same fields have shown that there exist populations of `hot' dust sources at comparable redshifts but with different IR spectral shapes (\citet{cha03a}, see also \citet{kha05}). Our higher redshift sources could be the lower-z tail of this radio-selected, 'hot dust' population.

\subsubsection{24$\mu$m selected samples}\label{comp24micron}
Most previous samples selected from $Spitzer$ surveys for IRS followup use a mid-infrared selection based on 24$\mu$m flux. It is important, therefore, to understand whether our 70$\mu$m sample differs from samples selected at 24$\mu$m. To make these comparisons, we combine our sample with that of \citet{brn08} for a total of 27 70$\mu$m selected sources, as the sample selections are complementary; our sample reaches fainter 70$\mu$m fluxes, but the optical cut is similar.

\subsubsubsection{Bright samples}
We first compare the mid-IR continuum properties of the combined 70$\mu$m sample to sources with high 24$\mu$m fluxes via the flux limited, f$_{24} >$ 10\,mJy sample in \citet{wee09}. To perform this comparison we use the rest-frame 15$\mu$m continuum luminosity and the $f_{24}/f_{15}$ continuum slope. If both these rest-frame wavelengths are seen in the IRS spectra then we measured these quantities directly; otherwise, we estimated fluxes at one or both wavelengths  via interpolation from a power law with a slope determined from comparison of observed f$_{\nu}$(24$\mu$m) and f$_{\nu}$(70$\mu$m). These interpolations should be regarded with caution, as they are sensitive to PAH and silicate contamination of the (observed-frame) 24$\mu$m band, which are difficult to compute for our sample as we either lack Long-Low data, or it is of relatively low signal-to-noise. The $\sim$20\% uncertainty assigned in Table 3 to these interpolated values for the continuum slope reflects the possibility that the observed 24$\mu$m flux density may not be purely a measure of dust continuum emission.

\begin{figure} 
\includegraphics[angle=90,width=170mm]{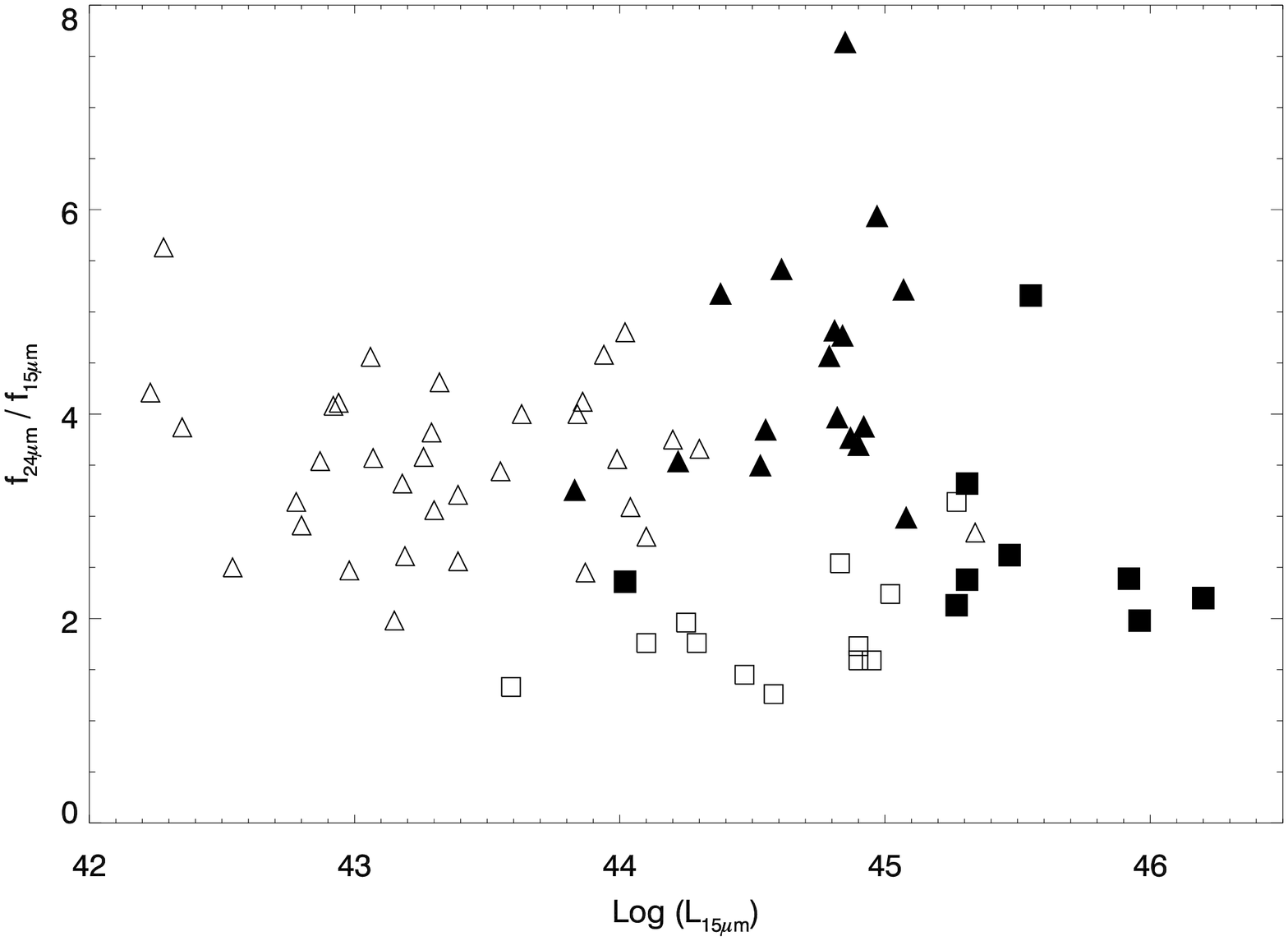}
\caption{
Distribution of continuum luminosity ($\nu$L$_{\nu}$(15$\mu$m) vs. 
rest rame continuum slope (f$_{\nu}$(24$\mu$m)/f$_{\nu}$(15$\mu$m)). Error bars have been omitted for clarity, but are in most cases comparable to the symbol sizes. Filled symbols are sources from our combined 70$\mu$m sample (Table 3 and \citet{brn08}), while open symbols are sources from the flux limited, f$_{\nu}$(24$\mu$m) $>$ 10\,mJy sample of \citet{wee09}. {\it Squares}: sources without detectable PAHs.  {\it Triangles}: PAH dominated systems. 
\label{lumslope}
 } 
\end{figure}

The comparison is shown in Figure \ref{lumslope}. The continuum luminosities for the 70$\mu$m sample are much greater than for the 24$\mu$m sample. The median log[$\nu$L$_{\nu}$(15$\mu$m)] (ergs s$^{-1}$) for the 70$\mu$m sample is 44.8 compared to 43.3 for the 24$\mu$m sample. This is straightforward to understand. The fainter optical and 24$\mu$m fluxes used for the 70$\mu$m selection allow the discovery of IR-luminous sources to much higher redshifts so we may reasonably expect to see more luminous sources. 
 
Interestingly, however, the luminosity differences between the samples may not be as large for the sources with weak PAH features. For these sources, the median log($\nu$L$_{\nu}$(15$\mu$m)) (ergs s$^{-1}$) for the 70$\mu$m sample is 45.4 compared to 45.0 for the 24$\mu$m sample, and the most luminous sources in both samples are similar, log($\nu$L$_{\nu}$(15$\mu$m)) $\simeq$ 46.2.  This comparison is not robust, given that the weak PAH sources in the 70$\mu$m sample only have the silicate feature in absorption, whereas the 24$\mu$m sample contains sources with the silicate feature in both absorption and emission.  Nevertheless, it seems that the fainter optical and 24$\mu$m fluxes used for the 70$\mu$m selection do not result in discovering more luminous {\it absorbed} sources, even though the sources are systematically at higher redshift.  We therefore suggest, with some reserve, that sources with weak PAHs and strong silicate absorption show weaker luminosity evolution with redshift than do PAH dominated sources. 

Considering the continuum slopes, we see results that would be expected from the 70$\mu$m selection; selecting sources at the longer wavelength favors sources with steeper spectra. For PAH sources, the median rest-frame ratio $f_{24}/f_{15}$ for the 70$\mu$m sample is 4.5 compared to 3.5 for the 24$\mu$m sample. For absorbed sources, the median rest-frame ratio $f_{24}/f_{15}$ for the 70$\mu$m sample is 2.7, compared to 1.7 for the 12 sources in the 24$\mu$m sample at sufficiently low redshifts to have a measurement.  For both PAH dominated sources and absorbed sources, the most extreme ratios are within the 70$\mu$m sample.   
 
These results demonstrate a systematic difference in effective dust temperatures for the 70$\mu$m sample compared to the 24$\mu$m sample.  For the 70$\mu$m sample, the steeper spectrum at rest frame $\sim$ 24$\mu$m implies a larger dust fraction at intermediate temperatures of $\sim$100 K.  It is also notable that the PAH dominated spectra are consistently steeper than the absorbed spectra in both samples. In the 70$\mu$m sample, the ratio $f_{24}/f_{15}$ is 4.5 and 2.7, respectively, and in the 24$\mu$m sample they are 3.5 and 1.7. This implies that the intermediate dust temperature component is more prominent in PAH dominated sources than in absorbed sources. For absorbed soures, if they contain a luminous AGN, the spectra can be flattened by having a more significant hot dust component to increase continuum emissivity at shorter wavelengths.   
 
 \subsubsubsection{Faint samples}
Finally, we compare our combined sample to those sources that are faint at observed frame 24$\mu$m ($f_{24}\lesssim2$mJy). This is a difficult comparison to make as the IRS spectra of these sources are usually of low signal to noise, making detailed comparisons difficult. We therefore make two adjustments. First, as most of our sample show PAH features, we restrict the comparison to those 24$\mu$m sources that also show PAH features by using the compilation in \citet{wee08}. This compilation includes new spectral measurements of faint sources, and published data from various IRS observing programs \citep{hou07, brn08, wee06a,far08, hou,wee06,yan07,pop08}. Second, we use a simple diagnostic that can be used for faint sources at a variety of redshifts -  the peak luminosity of the 7.7$\mu$m PAH feature\footnote{As we are using the peak luminosity of this feature, rather than its integrated flux, our luminosities differ from those quoted in  \citet{yan07} and \citet{pop08}} - and study how this peak luminosity evolves with redshift. The flux densities f$_{\nu}$(7.7$\mu$m) and luminosities $\nu$L$_{\nu}$(7.7$\mu$m) for our sample are in Table \ref{pahsils}, or in \citet{brn08}, while those for the faint 24$\mu$m samples are in \citet{wee08}.

\begin{figure}  
\includegraphics[angle=90,width=170mm]{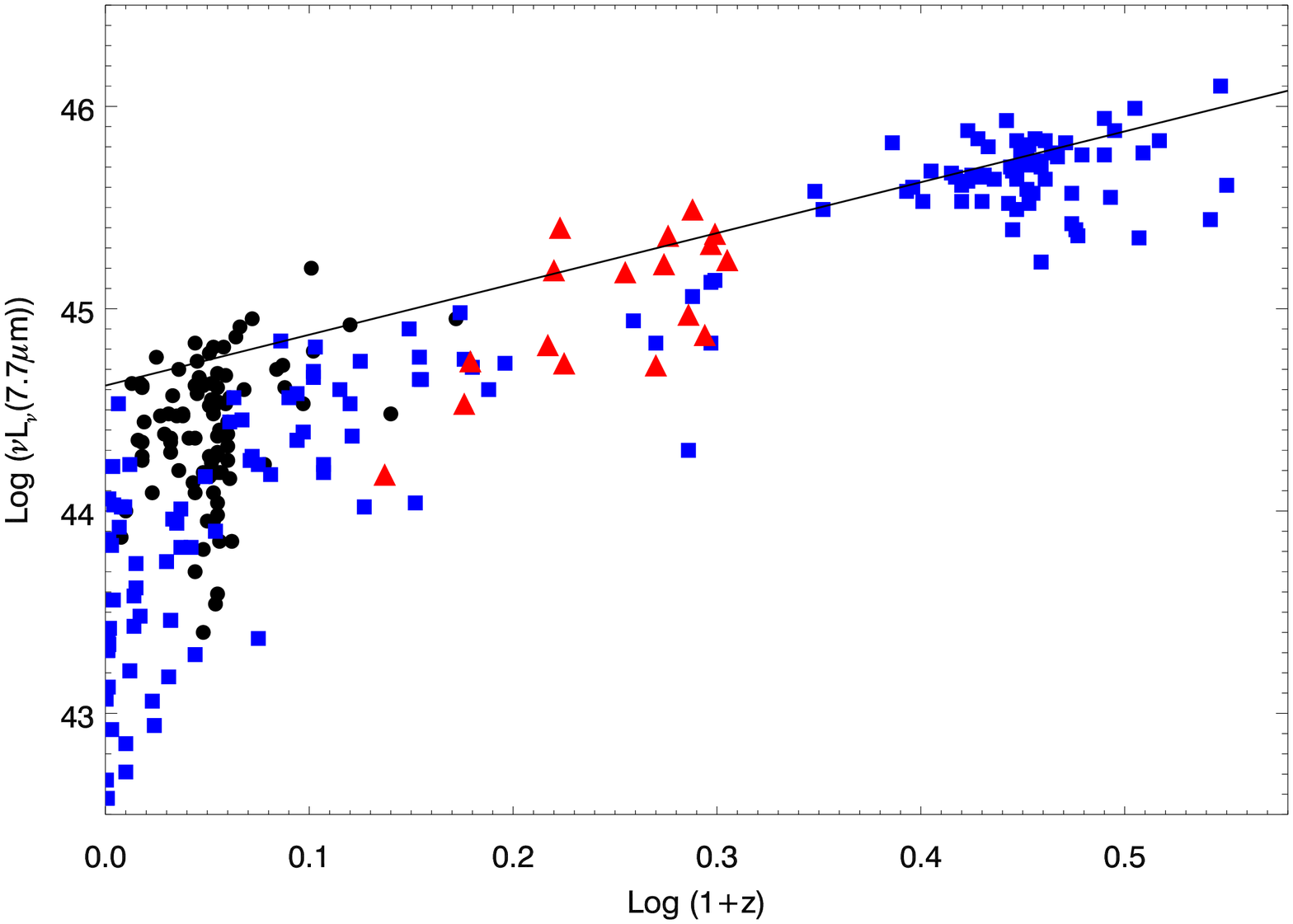} 
\caption{Distribution of PAH luminosity $\nu$L$_{\nu}$(7.7$\mu$m) with redshift. Error bars have been omitted for clarity, but are in most cases comparable to the symbol sizes. 
{\it Red Triangles}: PAH dominated sources from the combined 70$\mu$m sample from Table 3 and \citet{brn08}.
{\it Black Circles}: Starburst component of low redshift ULIRGs \citep{spo07,ima07}. 
{\it Blue Squares}: Starbursts in \citet{wee08}.
{\it Solid~line}: luminosity evolution to $z = 2.5$ of the form $(1+z)^{2.5}$, determined in \citet{wee08}. Results show that 70$\mu$m PAH sources are among the most luminous starbursts in the Universe, and fill a crucial redshift regime.
\label{pah70}
} 
\end{figure}

We plot these 7.7$\mu$m PAH luminosities against redshift in Figure \ref{pah70}. We also include the 24$\mu$m bright sources from \citet{wee08}, and low redshift ULIRGs from \citet{spo07}. Two important results can be seen. First, the 70$\mu$m selected starbursts are among the most luminous known from {\it any} infrared selected samples. They are more luminous, on average, than both low redshift ULIRGs or the \citet{yan07} sources at comparable redshifts, and approach the luminosities of the 24$\mu$m and sub-mm selected sources at $z>1.5$. This is expected - the high redshifts of our sample mean we are probing a greater volume and hence can find more luminous sources than local examples, and the additional demand of a 70$\mu$m detection means our sources will be more luminous, on average, than sources with just a 24$\mu$m detection at similar redshifts. 

Second, they reside in a redshift range, $0.5 \lesssim z \lesssim 1.5$, where other selection methods turn up relatively few sources. The 
faint 24$\mu$m samples, which have comparable 24$\mu$m fluxes to our sample but are not detected at $70\mu$m, span a significantly broader redshift range of $0.5\lesssim z \lesssim 3$, with the majority lying at $z>1.5$. It seems therefore that a 70$\mu$m selection of $\sim$20mJy, together with a faint optical and 24$\mu$m flux cut, serves to select sources almost entirely in the crucial redshift range $0.5 \lesssim z \lesssim 1.5$. This is the redshift range, for example, in which the results of \citet{lef} show steeper evolution of the IR-luminous galaxy population than is shown in Figure \ref{pah70}. We conclude that a 70$\mu$m selection, in combination with selections at mid-IR and far-IR/sub-mm wavelengths, is vital to measure adequately the luminosity evolution of luminous starburst galaxies over $0<z<2.5$.

\section{Summary}
Sixteen spectra have been obtained with the $Spitzer$ IRS of extragalactic sources in the SWIRE Lockman Hole field having f$_{\nu}$(70\ums) $>$ 19 mJy, including 12 sources with optical magnitudes $m_{r}>23$.   Results are combined with the sample of 11 sources with f$_{\nu}$(70\ums) $>$ 30 mJy from the NOAO Deep Wide-Field Survey region in Bootes \citep{brn08} to consider the nature of the 70$\mu$m population.

The 70$\mu$m sources are characterised either by strong PAH features or by strong silicate absorption features with weak or absent PAHs.  Ten of the 16 objects show prominent PAHs; four show strong silicate absorption, and two have no discernable spectral features. The continuum luminosities (measured by $\nu$L$_{\nu}$(15$\mu$m) in ergs s$^{-1}$) span 43.8 $<$ log $\nu$L$_{\nu}$(15$\mu$m) $<$ 46.3 with the absorbed sources having higher luminosities than the PAH dominated sources.  Compared to sources that are bright at 24$\mu$m  (f$_{\nu}$(24\ums) $>$ 10\,mJy), the 70$\mu$m sources have steeper rest frame mid-IR continua and higher luminosities. 

The 70$\mu$m sources with strong PAH features are among the most luminous starbursts seen at any redshift. Furthermore, these sources effectively fill the redshift range 0.5 $<$ z $<$ 1.5 where previous selection methods using a  24$\mu$m flux of $\sim1$mJy but without a 70$\mu$m detection have found few sources. This result demonstrates that selection of sources at 70$\mu$m to fainter flux limits will provide crucial samples for determining the evolution of star formation with redshift.

\acknowledgments
We thank the referee for a very helpful report. This work is based on observations made with the Spitzer Space Telescope, which is operated by the Jet Propulsion Laboratory, California Institute of Technology under NASA contract 1407. Support for this work by the IRS GTO team at Cornell University was provided by NASA through Contract Number 1257184 issued by JPL/Caltech. Support for the Spitzer Space Telescope Legacy Science Program, was provided by NASA through an award issued by the Jet Propulsion Laboratory, California Institute of Technology under NASA contract 1407. The research described in this paper was carried out, in part, by the Jet Propulsion Laboratory, California Institute of Technology, and was sponsored by the National Aeronautics and Space Administration. This research has made use of the NASA/IPAC Extragalactic Database (NED) which is operated by the Jet Propulsion Laboratory, California Institute of Technology, under contract with the National Aeronautics and Space Administration. DF thanks the Science and Technologies Facilities Council for support via an Advanced Fellowship.

\begin{deluxetable}{lcccc} 
\tablecolumns{5} 
\tablewidth{0pc} 
\tablecaption{Observations Log \label{obslog}} 
\tablehead{
\colhead{ID}&\colhead{Object}&\colhead{Modules\tablenotemark{a}}&\colhead{Exposure times}&\colhead{AOR Key\tablenotemark{b}} \\
\colhead{  }&\colhead{           }&\colhead{                                               }&\colhead{(s)}&\colhead{      }
}
\startdata 
 1  & SWIRE4 J103637.18+584217.0 & SL1/LL2/LL1 & 240/480/480 & 17410304 \\
 2  & SWIRE4 J103752.14+575048.6 & SL1/LL2/LL1 & 120/240/240 & 17413888 \\
 3  & SWIRE4 J103946.28+582750.7 & SL1/LL2     & 300/720     & 17413120 \\
 4  & SWIRE4 J104057.84+565238.9 & SL1/LL2     & 300/720     & 17411840 \\
 5  & SWIRE4 J104117.93+595822.9 & SL1/LL2     & 240/480     & 17410560 \\
 6  & SWIRE4 J104439.45+582958.5 & SL1/LL2     & 240/600     & 17412608 \\
 7  & SWIRE4 J104547.09+594251.5 & SL1/LL2/LL1 & 240/600/600 & 17413376 \\
 8  & SWIRE4 J104827.68+575623.0 & SL1/LL2     & 300/720     & 17410048 \\
 9  & SWIRE4 J104830.58+591810.2 & SL1/LL2     & 240/600     & 17413652 \\
 10 & SWIRE4 J104847.15+572337.6 & SL1/LL2/LL1 & 180/360/360 & 17412864 \\
 11 & SWIRE4 J105252.90+562135.4 & SL1/LL2/LL1 & 240/600/600 & 17410816 \\
 12 & SWIRE4 J105404.32+563845.6 & SL1/LL2/LL1 & 120/240/240 & 17412352 \\
 13 & SWIRE4 J105432.71+575245.6 & SL1/LL2     & 300/720     & 17409792 \\
 14 & SWIRE4 J105509.00+584934.3 & SL1/LL2     & 300/480/720 & 17412096 \\
 15 & SWIRE4 J105840.62+582124.7 & SL1/LL2/LL1 & 240/600/600 & 17414144 \\
 16 & SWIRE4 J105943.83+572524.9 & SL1/LL2/LL1 & 240/600/600 & 17411328 \\
\enddata
\tablenotetext{a}{Modules of the $Spitzer$ Infrared Spectrograph used to observe source.}
\tablenotetext{b}{Astronomical Observation Request number. Further details can be found by referencing these numbers within the {\it Leopard} software, available from the Spitzer Science Center.}
\end{deluxetable}

\begin{deluxetable}{lcccccccccc} 
\tablecolumns{11} 
\tablewidth{0pc} 
\tablecaption{Basic Data \label{sample}} 
\tablehead{
\colhead{ID}&
\colhead{z$_{phot}$\tablenotemark{a}}&
\colhead{z$_{irs}$\tablenotemark{b}}&
\colhead{m$_{r}$}&
\multicolumn{4}{c}{IRAC Fluxes ($\mu$Jy)}&
\multicolumn{3}{c}{MIPS Fluxes (mJy)} \\
\colhead{ }&
\colhead{ }&
\colhead{ }&
\colhead{ }&
\colhead{3.6$\mu$m}&
\colhead{4.5$\mu$m}&
\colhead{5.8$\mu$m}&
\colhead{8$\mu$m}&
\colhead{24$\mu$m}&
\colhead{70$\mu$m}&
\colhead{160$\mu$m}
}
\startdata 
 1  & \nodata & 0.97   & \nodata &  21.4 &  25.7 &  48.8   &  197.1  & 2.17 & 34.7  & \nodata  \\
 2  & (0.47)  & 1.55:  & 24.06   &  86.4 & 248.2 & 533.8   & 1248.2  & 3.98 & 19.9  & \nodata  \\
 3  & 1.30    & 0.90   & 23.36   &  28.0 &  25.2 & \nodata & \nodata & 1.05 & 22.8  & \nodata  \\
 4  & \nodata & 0.93   & 22.53   &  79.0 &  57.6 &  47.1   & \nodata & 1.05 & 24.2  & \nodata  \\
 5  & \nodata & 0.65   & 20.83   &  88.9 &  77.8 &  92.4   &  123.5  & 1.46 & 32.9  & 77.4     \\
 6  & \nodata & 0.68   & 21.49   &  46.1 &  49.5 &  59.0   &  139.1  & 1.21 & 22.0  & \nodata  \\
 7  & 0.52    & 0.39   & 20.21   & 115.9 & 105.5 & 129.7   &  319.3  & 1.77 & 20.2  & \nodata  \\
 8  & 1.33    & 0.86:: & \nodata &  40.7 &  41.0 &  33.3   &  138.0  & 0.98 & 37.4  & \nodata  \\
 9  & 1.14    & 0.94   & 23.28   & 146.9 & 122.8 & 114.4   &  114.4  & 1.62 & 20.6  & 125.0    \\
 10 & (1.22)  & 1.47:  & 23.74   & 154.0 & 286.1 & 455.7   &  799.1  & 2.62 & 23.2  & 124.0    \\
 11 & \nodata & 1.29:: & 20.39   & 236.6 & 165.1 & 164.9   &  176.2  & 1.53 & 30.2  & 71.3     \\
 12 & \nodata & 1.72:  & 23.29   &  23.4 &  27.2 & \nodata &  102.0  & 4.25 & 26.4  & \nodata  \\
 13 & \nodata & 1.02   & 23.29   &  72.4 &  66.6 &  64.9   &  113.0  & 1.18 & 37.0  & 75.9     \\
 14 & \nodata & 0.88   & \nodata &  68.4 &  63.9 &  67.9   &  120.4  & 0.97 & 24.1  & \nodata  \\
 15 & 1.06    & 0.89   & 23.85   & 112.3 &  83.6 &  89.4   &   92.6  & 1.50 & 19.3  & \nodata  \\
 16 & 1.18    & 0.80   & 23.16   & 170.8 & 180.0 & 218.2   &  310.3  & 1.96 & 30.7  & 119.8    \\
\enddata
\tablecomments{IRAC fluxes have errors of $5\%$; MIPS 24$\mu$m fluxes have errors of $10\%$, and MIPS 70$\mu$m and 160$\mu$m fluxes have errors of $20\%$.}
\tablenotetext{a}{Photometric redshift, derived using the code of \citet{rr07}. Redshifts in brackets are based on two optical bands and are not considered reliable.}
\tablenotetext{b}{Redshift derived from the IRS spectrum. These are accurate to $\Delta$z = 0.02, except sources noted by `:', which are accurate to $\Delta$z $\simeq$ 0.2, and those noted by`::' which have $\Delta$ z $\simeq$ 0.4 and should be regarded with caution.}
\end{deluxetable}

\begin{deluxetable}{ccccccccccccc}
\rotate
\tabletypesize{\scriptsize}
\tablecolumns{13}
\tablewidth{0pc}
\tablecaption{Spectral Measurements\label{pahsils}}
\tablehead{
\colhead{ID}&
\multicolumn{2}{c}{PAH 6.2$\mu$m}&
\colhead{f$_{7.7}$\tablenotemark{a}}&
\colhead{$\nu$L$_{7.7}$\tablenotemark{b}}&
\multicolumn{2}{c}{PAH 11.2$\mu$m}&
\colhead{$S_{sil}$}&
\colhead{f$_{24}$/f$_{15}$\tablenotemark{c}}&
\colhead{f$_{15}$\tablenotemark{d}}&
\colhead{log[$\nu$L$_{15}$]\tablenotemark{e}}&
\colhead{SFR\tablenotemark{f}}&
\colhead{L$_{IR}$\tablenotemark{g}}\\
\colhead{  }&
\colhead{Flux}&
\colhead{EW}&
\colhead{mJy}&
\colhead{ergs s$^{-1}$}&
\colhead{Flux}&
\colhead{EW}&
\colhead{}&
\colhead{}&
\colhead{mJy}&
\colhead{ergs s$^{-1}$}&
\colhead{M$_{\odot}$ yr$^{-1}$}&
\colhead{log(L$_{\odot}$)}
}
\startdata  
 1  & 4.20$\pm$0.65 & 0.18$\pm$0.06 & $<$3.1  & $<$45.41 & $<$2.0        & $<$0.3         & 2.30$\pm$1.00 & (2.4)   & 4.8     & 45.31   & $<$690  & 12.41 \\
 2  & $<$1.10       & $<$0.15       & $<$5.2  & $<$46.10 & $<$4.0        & $<$0.2         & 0.39$\pm$0.11 & (2.4)   & 6.8     & 45.92   & $<$3400 & 13.01 \\
 3  & 1.55$\pm$0.41 & 0.15$\pm$0.10 & 0.75    & 44.79    & \nodata       & \nodata        & $<$1.61       & (3.77)  & (1.76)  & 44.87   & 170     & 12.23 \\
 4  & 4.31$\pm$0.80 & 0.82$\pm$0.22 & 1.05    & 44.97    & \nodata       & \nodata        & $<$2.00       & (3.88)  & (1.86)  & 44.92   & 250     & 12.28 \\
 5  & 4.29$\pm$0.83 & 0.47$\pm$0.15 & 1.55    & 44.82    & $<$4.5        & $<$0.5         & 0.87$\pm$0.80 & (3.85)  & (1.63)  & 44.55   & 180     & 12.06 \\
 6  & 2.80$\pm$0.60 & 0.31$\pm$0.20 & 1.15    & 44.73    & 3.78$\pm$0.81 & 0.68$\pm$0.35 & 1.18$\pm$0.90 & (3.50)  & (1.41)  & 44.53   & 140     & 12.02 \\
 7  & 1.79$\pm$0.25 & 0.17$\pm$0.08 & $<$1.1  & $<$44.21 & 4.10$\pm$1.00 & 0.55$\pm$0.25 & 1.25$\pm$0.70 & 2.4     & 1.4     & 44.02   & $<$44   & 11.48 \\
 8  & \nodata       & \nodata        & 0.7     & 44.72    & \nodata       & \nodata        & \nodata       & (4.82)  & (1.68)  & 44.81   & 140     & 12.30 \\
 9  & 9.04$\pm$1.18 & 1.00$\pm$0.22 & 3.4     & 45.49    & \nodata       & \nodata        & \nodata       & (2.99)  & (2.61)  & 45.08   & 830     & 12.55 \\
 10 & $<$6.00       & $<$0.50       & $<$4.7  & $<$46.01 & $<$7.0        & $<$0.5         & 0.81$\pm$0.25 & (5.2)   & 3.3     & 45.55   & $<$2700 & 13.15 \\
 11 & \nodata       & \nodata        & \nodata & \nodata  & \nodata       & \nodata        & \nodata       & \nodata & \nodata & \nodata & \nodata & 12.80 \\
 12 & $<$5.60       & $<$0.30       & $<$6.8  & $<$46.30 & $<$8.5        & $<$0.76        & 1.32$\pm$0.24 & (2.2)   & (10.7)  & 46.20   & $<$5400 & 13.11 \\
 13 & 3.62$\pm$0.38 & 0.58$\pm$0.09 & 1.65    & 45.24    & \nodata       & \nodata        & \nodata       & (5.22)  & (2.17)  & 45.07   & 470     & 12.50 \\
 14 & 1.57$\pm$0.45 & 0.35$\pm$0.14 & 2.1     & 45.22    & \nodata       & \nodata        & 0.80$\pm$0.60 & (3.97)  & (1.63)  & 44.82   & 450     & 12.35 \\
 15 & 7.19$\pm$0.31 & 1.27$\pm$0.18 & 2.8     & 45.36    & $<$9.00       & $<$2.3        & 0.69$\pm$0.35 & (4.57)  & 1.50    & 44.79   & 620     & 12.36 \\
 16 & 5.20$\pm$0.81 & 0.36$\pm$0.10 & 2.3     & 45.18    & 4.79$\pm$1.66 & 1.24$\pm$1.00 & 1.94$\pm$1.00 & (3.70)  & 2.40    & 44.90   & 410     & 12.48 \\
\enddata  
\tablecomments{For the 6.2$\mu$m and 11.2$\mu$m PAH features, fluxes are in units of $10^{-22}$W cm$^{-2}$ and rest frame equivalent widths in $\mu$m.}
\tablenotetext{a}{Observed frame flux density at the peak of the 7.7$\mu$m PAH feature. Error is approximately 10$\%$.}  
\tablenotetext{b}{Rest frame PAH luminosity determined from the peak f$_{\nu}$(7.7$\mu$m).} 
\tablenotetext{c}{Rest-frame f$_{\nu}$(24$\mu$m)/f$_{\nu}$(15$\mu$m) continuum slope. Measurements are made from the IRS spectra if both rest-frame wavelengths are seen, and have a $\sim10\%$ error; otherwise, fluxes at one or both wavelengths are estimated via interpolation from a power law with a slope determined from comparison of observed f$_{\nu}$(24$\mu$m) and f$_{\nu}$(70$\mu$m). Ratios determined using such interpolations are in parentheses, and have errors of $\sim20\%$.} 
\tablenotetext{d}{Rest-frame 15$\mu$m flux density. Sources for which this is interpolated via a power law are in parentheses.}  
\tablenotetext{e}{Rest frame 15$\mu$m continuum luminosity $\nu$L$_{\nu}$(15$\mu$m). Sources for which this luminosity is interpolated via a power law are in parentheses.} 
\tablenotetext{f}{Star formation rate, determined from Equation \ref{sfreq}.}
\tablenotetext{g}{Rest-frame 1-1000$\mu$m luminosity derived from the SED fits described in \S\ref{sedfits}. The error on the luminosities is approximately 25\% in all cases, and does not include errors arising from uncertainties in redshift.}
\end{deluxetable}

\end{document}